\documentclass{article}
\usepackage{ijcai22}

\usepackage{times}
\usepackage{soul}
\usepackage{url}
\usepackage[hidelinks]{hyperref}
\usepackage[utf8]{inputenc}
\usepackage[small]{caption}
\usepackage{graphicx}
\usepackage{amsmath}
\usepackage{amsthm}
\usepackage{booktabs}
\usepackage{algorithm}
\usepackage{algorithmic}
\usepackage{amssymb}
\usepackage{xcolor}
\usepackage[caption=false]{subfig}
\title{Evolutionary Approach to Security Games with Signaling}

\author{
Adam {\.Z}ychowski$^1$\and
Jacek Ma{\'n}dziuk$^1$\footnote{Contact Author. Accepted at the 31st International Joint Conference on Artificial Intelligence (IJCAI 2022), Vienna, Austria}\and
Elizabeth Bondi$^2$\and
Aravind Venugopal$^3$\and
Milind Tambe$^2$\And
Balaraman Ravindran$^{3,4}$
\affiliations
$^1$Faculty of Mathematics and Information Science, Warsaw University of Technology\\
$^2$Center for Research on Computation and Society, Harvard University\\
$^3$ Robert Bosch Centre for Data Science and AI, IIT Madras\\
$^4$Department of Computer Science and Engineering, IIT Madras\\
\emails
\{a.zychowski, mandziuk\}@mini.pw.edu.pl,
ebondi@g.harvard.edu,\\
aravindvenugopal19@gmail.com,
milind\_tambe@harvard.edu,
ravi@cse.iitm.ac.in
}

\begin{document}

\maketitle

\begin{abstract}
Green Security Games have become a popular way
to model scenarios involving the protection of natural resources, such as wildlife. Sensors (e.g. drones equipped with cameras) have also begun to play a role in these scenarios by providing real-time information. %, often from multiple locations simultaneously. 
Incorporating both human and sensor defender resources strategically is the  subject of recent work on Security Games with Signaling (SGS). However, current methods to solve SGS do not scale well in terms of time or memory. %, particularly in certain cases. 
We therefore propose a novel approach to SGS, which, for the first time in this domain, employs an Evolutionary Computation paradigm: EASGS. EASGS effectively searches the huge SGS solution space via suitable solution encoding in a chromosome and a specially-designed set of operators. The operators include three types of mutations, each focusing on a particular aspect of the SGS solution, optimized crossover and a local coverage improvement scheme (a memetic aspect of EASGS). We also introduce a new set of benchmark games, based on dense or locally-dense graphs that reflect real-world SGS settings. % and perform a comprehensive EASGS verification on 342 game instances. 
In the majority of 342 test game instances, EASGS outperforms state-of-the-art methods, including a reinforcement learning method, in terms of time scalability, nearly constant memory utilization, and quality of the returned defender's strategies (expected payoffs). %The results demonstrate the efficacy and robustness of the method, especially when applied to games played on dense or locally-dense graphs which reflect real-world SGS settings. 
%Due to near constant memory scalability and \emph{anytime} characteristics,
%EASGS can be employed to arbitrarily
%large games as opposed to state-of-the-art methods,
%which are limited by either memory or computation
%time requirements.
\end{abstract}

\section{Introduction}
\label{sec:introduction}
% In recent years it can be observed a tendency to design computer programs which helps people making better decisions in many real-life scenarios. One of such areas is security management and public surveillance. In this domain, problems are often modeled as Stackelberg Security Games with two competing players. One of the players - the \emph{defender} commits to a strategy first and then the opponent (attacker) chooses their strategy - based on the already announced leader's strategy. The above concept fits a wide range of real-life security-related scenarios in which the attacker can observe the guards (defender) and discover their patrolling strategy. Finding Stackelberg Equilibrium~\cite{leitmann1978generalized} in such games allows the defender to setup an optimal strategy that minimizes losses related to attack consequences.

% Systems based on Stackelberg Security Games were already successfully applied to a wide range of real-world scenarios, for instance scheduling Los Angeles International Airport canine patrols~\cite{jain2010software}, fare inspections in Los Angeles Metro system or prevent poaching~\cite{yin2012trusts}, US Coast Guard's resources in Boston harbour~\cite{shieh2012protect}, and protecting wildlife in Queen Elizabeth National Park in
% Uganda~\cite{yang2014adaptive}. Please refer to~\cite{sinha2018stackelberg} for many other examples.

% All of the hitherto proposed methods for solving SSG employ Mixed-Integer Linear Programming (MILP). 

Artificial intelligence is increasingly being used as a decision aide in many real-world scenarios, %particularly those in which there are limited resources. One 
including for the protection of natural resources, such as wildlife, %forests, and fisheries, 
from illegal activities, such as wildlife poaching. %, logging, and fishing. 
These domains are often modeled as Green Security Games (GSGs) with two competing players \cite{fang2015security,wang2019deep}, the defender and adversary. %The defender player commits to a strategy first, and then the adversary chooses their strategy based on the defender’s. 
Finding the Stackelberg Equilibrium \cite{leitmann1978generalized} in such games allows the defender to design an optimal strategy that minimizes attacks given limited resources. These strategies have proven to be effective in multiple real-world deployments, including in Queen Elizabeth National Park in Uganda \cite{yang2014adaptive}.

Recently, sensors
%, including camera traps and conservation drones, 
have emerged for use in these domains as additional defender resources \cite{munoz2013introducing,basilico2014strategic,basilico2015security,de2018facing,bondi2018spot}. Because they have the potential to detect an adversary in real time, they hold promise in improving GSG strategies. However, attacks may still occur if no human defender (henceforth, patroller) is nearby to interdict. %that result from sensors further stretch the typically limited resources. 
Security games with signaling (SGS) have been proposed to help address this \cite{bondi2020signal}. In these games, warning signals are sent to an adversary to convey  that a patroller is responding, even sometimes when no patroller is truly responding. %Deceptive signals, where a patroller does not actually respond, allow us to protect areas without patrollers and therefore achieve better performance. 
These must be used strategically \cite{bondi2020signal,xu2018strategic}, as a total lack of response after a signal would render it meaningless. %By strategically signaling,  are able to achieve better performance than without sensors or signaling. %This methodology was considered in \cite{xu2018strategic,bondi2020signal}, in which wildlife conservation and conservation drones are primarily considered. may turn on a light aboard the drone when a human is detected, in hopes of deterring poaching.

However, these methods for solving SGS employ traditional optimization techniques, such as Linear Programming (LP) or Mixed Integer Linear Programming (MILP) and suffer from poor time scalability and large memory requirements. In this paper, we aim to solve SGS with Evolutionary Algorithms (EAs), which are inspired by the process of (biological) evolution. 
%{\color{violet}Although there are usually no mathematical proofs of convergence to the optimal solution for EA-based algorithms, due to their experimentally proven efficiency they have been widely used in a plethora of real-life optimization problems, e.g. ~\cite{coello2004applications,hu2015application}.}
Recently, EAs have been successfully applied to sequential security games~\cite{ZychowskiMandziuk2021}, including games played in continuous space with moving targets~\cite{karwowski2019memetic}, or games with bounded rationality~\cite{zychowski2021learning}.

%Recently, EAs have been successfully applied to sequential security games: \cite{ZychowskiMandziuk2021} introduced a generic framework for solving such games and \cite{karwowski2019memetic} proposed a memetic algorithm for games played on a continuous space with moving targets (e.g. ferries at sea).

%This work is the first to propose EAs for SGS.
%The vast majority of domain methods rely on MILP solutions which - by design - suffer from scalability limitations.
Motivated by finding a scalable algorithm for SGS, we propose Evolutionary Algorithm for Security Games with Signaling (EASGS), the first metaheuristic for SGS based on an EA. %, leading to certain improvement of results and offering better time scalability. %The proposed method is generic, 
%(and therefore widely applicable in the considered domain), 
%while also tuned to our particular problem thanks to the memetic local optimization component. 
% EASGS is tailored towards SGS by means of chromosome (solution) representation, a suitable population refreshing scheme, and a memetic optimization phase in the form of a local coverage improvement scheme. Furthermore, we use novel problem-optimized evolutionary operators: three types of mutation, each addressing a particular aspect of the SGS formulation, and a dedicated crossover procedure are proposed.
%We believe that this first application of a metaheuristic method to signaling games, leading to certain improvement of results and offering better time scalability, may lay the framework for wider application of metaheuristic methods to security games.
%
%
%Unfortunately, these previous methods are applied to different game models and cannot be directly used for SGSs. Secondly, EASG contains some novel ideas of evolutionary operators, for instance, new mutation types and population \textit{refresh}. Also, the whole fitness function evaluation process is different in order to address SG characteristics. 
%The proposed method is evaluated on a wide range of benchmark games and proven to excel the competitive approaches in terms of strategy quality (for the majority of test games), memory utilization, and time scalability.
%
%\subsubsection{Contribution}
%\smallskip
%\noindent
%\textbf{Contribution.}
In particular, we propose: % therefore as follows: %threefold:
(a) a novel %EA-based approach to SGS with an efficacious 
defender-centered chromosome representation to facilitate an EA-based approach to SGS,
(b) evolutionary operators, including a novel mutation and crossover approach, %consisting of three types of mutation, 
a population refreshing scheme, and local (memetic) optimizations, 
(c) a new set of benchmark games with various underlying graph structures, designed to reflect real-life SGS scenarios,
%(d) substantial modification of the reinforcement learning approach~\cite{venugopal2021reinforcement} in order to adapt the algorithm to the SGS problem and compare with the proposed EA-based method,
(d) improved performance over state-of-the-art methods based on MILP and reinforcement learning (based on a substantial adaptation of \cite{venugopal2021reinforcement}, which we call m-CombSGPO) in terms of expected payoffs, time scalability, and memory.

Due to space limits, certain aspects related to the game scenario, EASGS implementation details, parameterization, and benchmark games are further discussed in supplementary material.

\section{Problem Definition}\label{sec:problem-definition}

We consider repeated interactions between defender and adversary agents, assuming there is only one adversary. Note that EASGS can be easily applied to games with multiple adversaries, as only chromosome evaluation needs to be modified, however, we focus on one adversary for comparison purposes. The defender resources consist of $l$ sensors and $k$ patrollers. The sensors can detect an adversary in real time and notify the patrollers. There is uncertainty in the detection, as automatic detection in conservation drone imagery may be imperfect, particularly given occlusions such as trees. We specifically consider false negative detections as in \cite{bondi2020signal,basilico2016security,basilico2017adversarial}.
% Sensors may be conservation drones with thermal infrared cameras aboard and an automated detection system, as specifically considered in \cite{bondi2020signal}. 

In addition to sending notifications after a detection, sensors can also send two potential signals to an adversary, represented by $\sigma_0$ and $\sigma_1$. We consider $\sigma_0$ to be a weak signal, e.g., if considering a conservation drone as in \cite{bondi2020signal}, no lights are illuminated aboard the drone. On the other hand, $\sigma_1$ is a strong signal, e.g., lights are illuminated aboard the drone. Typically, we think of the strong signal as indicating that a patroller is (likely) responding, i.e., the lights turn on to deter the adversary. 
%The set of possible signaling states the adversary could observe is represented by $\Omega = \{n, \sigma_0, \sigma_1\}$
%, where $n$ means that there is no defender resource whatsoever. 
%There is no state for a patroller, as the adversary would be immediately captured. 
There is also observational uncertainty when the adversary observes the signal, again due to potential occlusions or difficulties viewing the true signal.

Sensors cannot directly prevent an attack, only signal to an adversary or notify the patrollers regarding an observation (which we assume is always done on an observation). If the adversary is deterred from attacking by a signal, both the defender and adversary receive 0 utility. If patrollers are at a target with an adversary, or nearby, they can interdict. The defender receives a positive utility if an attack is prevented successfully, while the defender receives a negative utility if an attack is successful. The opposite is true for the adversary. 

The geography of the protected region is modeled using a graph, $G=(V,E)$, where each potential target in the region is represented by a vertex $\nu \in V$. If a sensor and  patroller are at vertices connected by an edge $\epsilon \in E$, the patroller can successfully respond to that sensor’s notification to prevent an attack. The number of vertices, $|V|$, %in graph $G$ 
is denoted by $\mathcal{N}$.%=|V|$.

\begin{figure}[t]
\centering
    \includegraphics[width=1.0\columnwidth]{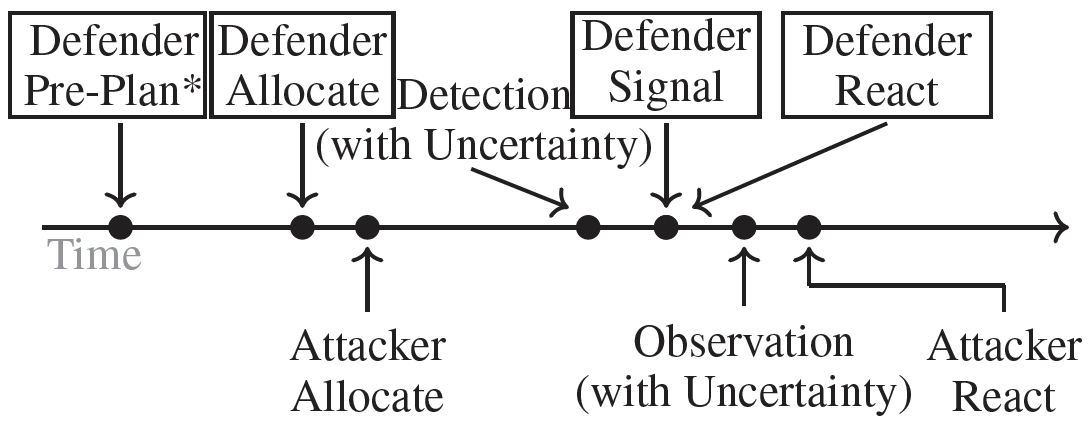}%
  \caption{Players' actions timeline~\protect\cite{bondi2020signal}.
  }
  \label{fig:game-timing}
\end{figure}

If there are no observations by either patrollers or sensors, patrollers may move to a neighboring vertex to have another chance of preventing an attack. This is known as the reaction stage, and is particularly useful if there is detection uncertainty, as a patroller would be able to move to a neighboring vertex with a sensor if the sensors have high uncertainty. Therefore, the order of the game is as follows: the defender first fixes their mixed strategy offline, then carries out a pure strategy allocation. The adversary proceeds to choose a target, and then may be detected by the defender's sensors. The defender may send a signal from the sensor and/or patrollers to respond to notifications, or re-allocate in the reaction stage. The adversary may observe the signal and react (flee). 
The game scenario is depicted in Figure~\ref{fig:game-timing}.

A defender pure strategy, $e$, therefore consists of an allocation of $l$ sensors, an initial allocation of $k$ patrollers, and a reaction stage allocation of $k$ patrollers to neighboring vertices. The defender also has a signaling scheme, which is a series of probabilities over all targets and states.
There are different signaling schemes when there is or is not a detection, as \cite{bondi2020signal} found that it was necessary to occasionally signal without a detection due to uncertainty. Per the Stackelberg model, the adversary knows (only) the defender mixed strategy and signaling scheme.

% in that \cite{venugopal2021reinforcement} works only for grid grawhich assumes the adversary knows the initial defender pure strategy before the defender resources begin dynamic patrolling over a grid, and is therefore not directly comparable.

In SGS, two sources of uncertainty are considered: \textit{detection uncertainty} and \textit{observational uncertainty}.
The \textit{detection uncertainty} is related to the limited capability of sensors, e.g., false negative detections. 
In real-life scenarios, a sensor's imperfection (whether the detection algorithm, the sensor itself, or both) can cause mistakes in reporting an adversary's presence. The most common situation is a false negative detection, i.e., a sensor misses the adversary due to, for example, poor visibility, weather conditions, occlusions, or even a power outage. In order to account for this, a probability of its occurrence is denoted by $\gamma$.
Namely, the sensor with a probability equal to $\gamma$ does not note the adversary's presence in the target during an attack.

The \textit{observational uncertainty} is related to the adversary's observation of a signal. The true state may differ from the adversary’s observation, i.e., the adversary may not notice the presence of a sensor or signal, possibly due to occlusions or other conditions. There are therefore 3 possible signaling states: (1) the sensor is not present in the target vertex at all (denoted by $n$), (2) the sensor is present, but does not send a strong signal (it sends a weak signal: $\sigma_0$), (3) the sensor is present and sends a strong signal: $\sigma_1$. Let us denote a conditional probability that the adversary will observe state $\hat{\omega}$ while the true signaling state is $\omega$ by $P[\hat{
\omega}|\omega]$. Then, we can define an observational uncertainty matrix $\Pi$.

$$\Pi = \begin{bmatrix} P[n|n] & P[n|\sigma_0] & P[n|\sigma_1] \\
P[\sigma_0|n] & P[\sigma_0|\sigma_0] & P[\sigma_0|\sigma_1] \\
P[\sigma_1|n] & P[\sigma_1|\sigma_0] & P[\sigma_1|\sigma_1]
\end{bmatrix}$$

\section{Related work}
\label{sec:state-of-the-art}

\cite{bondi2020signal} achieves state-of-the-art performance solving the problem described in the previous section. We omit the details for the sake of brevity, but in short, they adapt the multiple linear programs (LPs) approach from \cite{conitzer2006computing}. %In this approach, an LP is solved for each vertex, assuming that the adversary chose this vertex to attack. 
By maximizing the defender expected utility for an LP at each vertex, % under this assumption, 
the strategy yielding the highest defender expected utility of all LPs is the optimal strategy overall. Unfortunately, each LP is exponential in the number of possible pure strategies. The baseline method proposed in \cite{bondi2020signal} solves each exponential LP to get an exact solution. We will also compare with this when possible, and refer to it as the exponential LP method (SELP). 

In order to solve the problem more efficiently, \cite{bondi2020signal} adapt the branch-and-price method. The overall idea is that they can (1) speed up solving each individual LP by using column generation, i.e., adding pure strategies gradually based on the result of a mixed integer linear program (MILP), and (2) use the branch and bound technique by solving a relaxation of each LP and pruning.  %several of the LPs altogether based on the solutions. 
We will denote this algorithm as branch-and-price (SBP) in comparisons. The most efficient method proposed in \cite{bondi2020signal} uses the same branch-and-price method, but greedily generates an initial set of pure strategies for the LPs. We will denote this as branch-and-price with warm-up (SBP+W).

Another approach (called CombSGPO) proposed in~\cite{venugopal2021reinforcement} is based on reinforcement learning, using competitive gradient descent and a multi-agent Deep Q-Network~\cite{van2016deep} (MADQN) to solve a zero-sum game between the defender and the adversary on a grid graph.
Our game model differs from~\cite{venugopal2021reinforcement} by imposing no restrictions on the game graph topology, allowing a wider range of adversary's observation-reaction scenarios, and using different defender's strategy representation.
%which uses a signaling scheme where sensors either signal the attacker or notify the patroller, where the latter action cannot be observed by the attacker and the former always leads to the attacker fleeing, if observed by the attacker. CombSGPO works only on grid graphs and assumes that attackers have access to the initial allocations of defender resources, rather than the defender mixed strategy for allocation.} 
We modify CombSGPO to work with our general-sum game model on any type of graph and call it m-CombSGPO (\textbf{m}odified CombSGPO). 
m-CombSGPO uses an actor-critic algorithm for computing the defender mixed strategy for allocation and reallocation, in the space of defender pure strategies 
%used in 
\cite{bondi2020signal}. Observing this mixed strategy, the adversary, represented by an actor-critic algorithm, chooses the target. Since we have a reaction stage instead of a patrolling stage like in \cite{venugopal2021reinforcement}, instead of using a MADQN, we use a sensor signaling network, a neural network to choose the sensor (drone) signaling strategy, followed by an adversary decision neural network to choose whether the adversary should continue attacking or flee, both trained using the policy gradient algorithm \cite{sutton2000policy}. 
m-CombSGPO is used as a reinforcement learning baseline.

\section{Evolutionary Algorithm}
\label{sec:EASGS}

We propose a novel Evolutionary Algorithm for Security Games with Signaling (EASGS). %The backbone of the method is based on a well-established EAs framework. However, the strength of the proposed solution stems from carefully designed evolutionary operators (mutation and crossover) and local optimizations which address SGS specificity and difficulties.
EASGS is a population-based algorithm, which means that a set of individuals representing solutions (i.e., defender's mixed strategies) iteratively evolve to find the optimal solution. Each individual is represented in the form of a chromosome. %, encodes a valid solution, i.e., a defender's mixed strategy.
Initially, a population of $n_{pop}$ chromosomes is randomly generated. Then, evolutionary operators (crossover, mutation, local optimization, and selection), which are described in detail in the following subsections, are repeatedly applied in subsequent generations. A total of $n_{pop}$ chromosomes are maintained throughout the process of applying these operators, and ultimately, this leads to a final set of solutions. %During selection, the top $n_{pop}$ chromosomes are promoted to the next generation based on their fitness values, calculated as the defender's payoff when playing the mixed strategy encoded in the chromosome against an optimal adversary's response. 
%Throughout this process, the size of the population $n_{pop}$ remains constant, while the chromosomes in the population change after every generation. 

%\subsubsection{Chromosome representation}
\subsection{Chromosome representation}
Each chromosome encodes a valid solution, i.e., a candidate defender's mixed strategy, in the form of a list of pure strategies and their respective probabilities, and the sensors' signaling strategy:\\
$CH_j = \{(e^j_1,q^j_1), \ldots, (e^j_i,q^j_i), \ldots, (e^j_{d_j},q^j_{d_j}), \boldsymbol{\Psi^\theta_j}, \boldsymbol{\Phi^\theta_j}\}$\\ 
where $CH$ represents the chromosome, %\\
$j \in \{1,\ldots, n_{pop}\}$ is the individual's identifier, %\\
$d_j$ is the number of pure strategies included in the mixed strategy represented by the chromosome $CH_j$, %\\
$e^j_i \in \mathcal{E}$ is a pure strategy in the set of pure strategies, %\\
$q^j_i \in [0,1]$ is the probability of pure strategy $e^j_i$, $\sum_{i=1}^{d_j}q_i^j=1$. \\
%$\Psi^j = [\psi_1^j, \psi_2^j, \ldots, \psi_{\mathcal{N}}^j]$ is the sensors' signaling strategy in the case of an adversary detection, i.e., the sensor located in the vertex $i$ sends signal $\sigma_0$ with probability $\psi_i^j \in [0,1]$ when it detects an adversary, \\
$\theta \in \{\bar{s}$, $s^+$, $s^-$\} are allocation states: $\bar{s}$ when no patroller is in the sensor's neighbourhood (i.e., no one can respond to a notification), 
$s^+$ when a sensor has a patroller who will visit in the reaction stage (and could respond),
$s^-$ when no patroller will visit in the reaction stage (but there is a patroller in at least one neighbouring vertex who could respond),\\
$\boldsymbol{\Psi^\theta_j} = [\Psi^\theta_{j,1}, \Psi^\theta_{j,2}, \ldots, \Psi^\theta_{j,\mathcal{N}}]$ is the sensors' signaling strategy when an adversary is detected, i.e., $\Psi^\theta_{j,\nu}$ is the conditional probability\footnote{In \cite{bondi2020signal}, $\psi^{\theta}/x^{\theta}$, where $x$ is the marginal probability of state $\theta$.} that the sensor at vertex $\nu$ sends signal $\sigma_0$ given the allocation state $\theta$ and that it detects an adversary,\\
$\boldsymbol{\Phi^\theta_j} = [\Phi^\theta_{j,1}, \Phi^\theta_{j,2}, \ldots, \Phi^\theta_{j,\mathcal{N}}]$ is the sensors' signaling strategy in the case of no adversary detection, i.e., $\Phi^\theta_{j,\nu}$ is the conditional probability %\footnote{In \cite{bondi2020signal}, $\phi^{\theta}/x^{\theta}$.} 
that the sensor at vertex $\nu$ sends signal $\sigma_0$ given the allocation state $\theta$ and that it does \textit{not} detect an adversary.

%Please note that probabilities $\Psi^j$ and $\Phi^j$ are conditional probabilities - $\phi_i^j = Pr[$sensor is located in the vertex $i$ $|$ sensor sends signal $\sigma_0]$ and $1-\phi_i^j = Pr[$sensor is located in the vertex $i$ $|$ sensor sends signal $\sigma_1]$.

Pure strategies, $e$, are represented by tuples which contain defender's decisions according to the game rules described in problem definition section: $e = (V_p, V_s, V_r)$, where
$V_p = \{\nu^p_1, \nu^p_2, \ldots, \nu^p_k\} \subseteq V$ is the patrollers' allocation plan, i.e., a set of vertices in which patrollers will be placed,
$V_s = \{\nu^s_1, \nu^s_2, \ldots, \nu^s_l\} \subseteq V$ is the sensors' allocation plan, %\\
$V_r = \{\nu^r_1, \nu^r_2, \ldots, \nu^r_k\} \subseteq V$ is the patrollers' re-allocation plan, i.e., the neighboring vertices to which each patroller moves if no adversaries are observed; the $i_k$-th patroller moves from vertex $\nu^p_{i_k}$ to $\nu^r_{i_k}$ (if relevant edge in the graph exists).

Observe that in~\cite{bondi2020signal}, a pure strategy is defined in a vertex-centered way, i.e. each vertex is assigned an allocation state (please consult~\cite{bondi2020signal} for the details).
In EASGS, we propose a defender-centered strategy representation. Instead of focusing on the state of each vertex, a list of ``actions’’ to be performed by the defender is encoded (i.e., to which vertices patrollers/sensors will be allocated/reallocated). Both encodings are equivalent, albeit, the latter one makes formulation of evolutionary operators (especially mutation) simpler and more effective.

%\subsubsection{Initial population}
\subsection{Initial population}
At the beginning, the population is initialized with individuals containing only one pure strategy: $\forall_j\  d_j = 1\  \land\  q^j_1 = 1$. These strategies are generated randomly according to the game rules defined in the problem definition, i.e., $V_p, V_s, V_r$ are randomly selected subsets of $V$, and each element of $\boldsymbol{\Psi^\theta}$ and $\boldsymbol{\Phi^\theta}$ (signaling probability) is chosen uniformly at random from the unit interval.

\begin{figure}[ht]
\centering
    \includegraphics[width=\columnwidth]{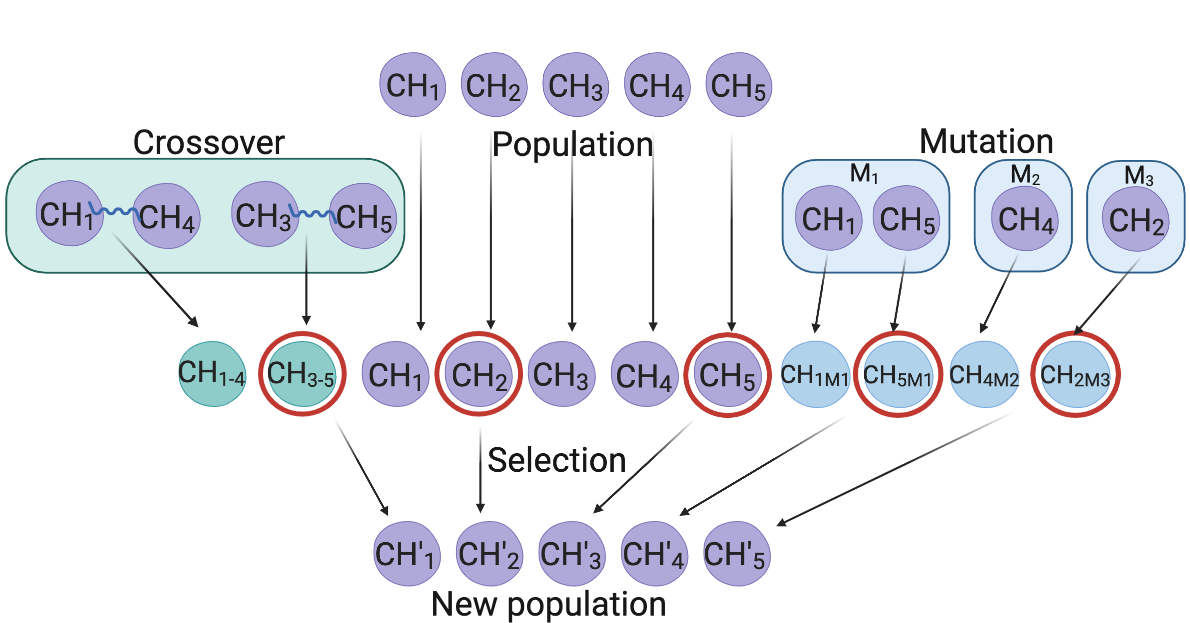}%
  \caption{An example of EASGS operators application. %Random subset of the population is selected for crossover and mutation. Then all chromosomes (original ones from the population and results of evolutionary operators) take part in selection process. At the end, a new population with the same number of chromosomes as the initial one is generated.
  }
  \label{fig:easgs_overview}
\end{figure}

%\subsubsection{Crossover}
\subsection{Crossover}
The crossover operation combines two chromosomes by merging their sets of pure strategies such that the resulting child strategy contains a union of two parent strategies. A probability of each pure strategy in the resulting child chromosome depends on the previous probability (in the parent chromosome) and its utility (defender's payoff in case of playing this pure strategy). Formally, the probability of pure strategy $e$ in the child chromosome is set to: $2^{u(e)} q$ where $q$ is the probability of strategy $e$ in the parent chromosome and $u(e)$ is its utility, i.e. the defender's payoff when playing strategy $e$, normalized to interval $[-1;1]$.  In particular, probability of the best pure strategy is doubled (its normalized utility equals $1$), whereas that of the pure strategy with the lowest utility is halved (normalized utility equals $-1$). After that, the probabilities of all pure strategies are normalized so as to sum to $1$. The intuition behind the above setting is to strengthen pure strategies with higher utilities and weaken lower-valued ones while, at the same time, not neglecting any of them. 
The above crossover definition extends the one proposed in~\cite{ZychowskiMandziuk2021} in which utilities of strategies are not considered and all probabilities are first halved, and then normalized to $1$.

Sensors' signaling strategies $\boldsymbol{\Psi^\theta}$ and $\boldsymbol{\Phi^\theta}$ in the child chromosome contain  the averaged signaling probability from each parent.  
A result of the crossover operation on chromosomes $CH_1$ and $CH_2$ 
 will be the new chromosome, $CH_{1\text{-}2}$:
\begin{multline*}
CH_{1\text{-}2}=\\
\{(e^1_1,2^{u(e^1_1)}q^1_1),(e^1_2,2^{u(e^1_2)}q^1_2), \ldots, (e^1_{d_1},2^{u(e^1_{d_1})}q^1_{d_1}),\\(e^2_1,2^{u(e^2_1)}q^2_1),(e^2_2,2^{u(e^2_2)}q^2_2), \ldots, (e^2_{d_2},2^{u(e^2_{d_2})}q^2_{d_2}),\\ \boldsymbol{\Psi^\theta_{1\text{-}2}}, \boldsymbol{\Phi^\theta_{1\text{-}2}}\},
\end{multline*}
where\\
%\smallskip
$\boldsymbol{\Psi^\theta_{1\text{-}2}} = [\frac{1}{2}(\Psi^\theta_{1,1} + \Psi^\theta_{2,1}), % \frac{1}{2}(\Psi^\theta_{1,2} + \Psi^\theta_{2,2}), 
\ldots,
\frac{1}{2}(\Psi^\theta_{1,\mathcal{N}} + \Psi^\theta_{2,\mathcal{N}})]$, \\
$\boldsymbol{\Phi^\theta_{1\text{-}2}} = [\frac{1}{2}(\Phi^\theta_{1,1} + \Phi^\theta_{2,1}), % \frac{1}{2}(\Phi^\theta_{1,2} + \Phi^\theta_{2,2}), 
\ldots,
\frac{1}{2}(\Phi^\theta_{1,\mathcal{N}} + \Phi^\theta_{2,\mathcal{N}})]$.
%\smallskip

Repeating crossover operation in subsequent generations without any means of strategy removing would lead to chromosomes with a large number of pure strategies with tiny probabilities. To avoid this, after the crossover operation, each pure strategy $e_i^j$ in the child chromosome $CH_j$ is deleted with probability $(1-q^j_i)^2$. The choice of deletion probability was empirically found to keep a good balance between the removal of less significant strategies (with lower $q^j_i$ values) and the overall size of a mixed strategy. 
%The removal procedure favors pure strategies with higher probability (presumed more important) and %preserves reasonable sizes of mixed strategies. 
After deletion, the probabilities of the remaining pure strategies are normalized so as to 
%ensure that their probabilities 
sum to $1$, i.e., $q^j_i\prime := \frac{q^j_i}{\Sigma_{\tau} q^j_{\tau}}$.

A set of chromosomes that are subject to crossover is chosen randomly in the following way. First, each individual from the population is selected for the crossover with crossover probability $\mathcal{P}_{c}$. Then, all selected chromosomes are randomly paired, and a crossover operation is performed. In the case of an odd number of individuals, a randomly chosen one is removed. From each set of parents, one new child chromosome is created and added to the population. An example crossover process is presented in Figure~\ref{fig:easgs_overview}. The crossover operation helps in exploitation of the solution space by mixing strategies found so far. % and potentially benefiting from their synergy.

%\subsubsection{Mutation}
%\label{sec:mutation}
\subsection{Mutation}
The mutation operator is applied to each chromosome $CH_j$ from the current population independently with probability $\mathcal{P}_{m}$. There are 3 types of mutations: \textit{probability change} ($M_1$), \textit{allocation/signaling strategy modification} ($M_2$) and \textit{coverage improvement} ($M_3$).

\textit{Probability change} ($M_1:\mathbb{R}^{d_j} \mapsto \mathbb{R}^{d_j}$) assigns a new probability $\rho$ (uniformly drawn from the unit interval) to a randomly selected pure strategy $e^j_i$. Then, a softmax function is applied to all probabilities in a chromosome: $\forall_i\  q^j_i\prime := \frac{q^j_i}{\Sigma_{\tau} q^j_{\tau}}$. Thus, the $M_1$ mutation can be formulated as follows: $M_1([q^j_1, \ldots, q^j_i, \ldots, q^j_{d_j}]) = [q^j_1\prime,\ldots, \rho, \ldots, q^j_{d_j}\prime]$.
$M_1$ modifies the influence of particular pure strategies on the result.

\textit{Allocation/signaling strategy modification} ($M^1_2: \mathcal{E} \mapsto \mathcal{E}$, $M^2_2: [0,1]^{\mathcal{N}}\mapsto [0,1]^{\mathcal{N}}$) randomly selects one of the pure strategies ($M^1_2$) or signaling strategy ($M^2_2$) in the mutated chromosome. In case of $M^1_2$, a random perturbation is applied to a randomly chosen part of that pure strategy ($V_p, V_s$, or $V_r$). It means that the randomly selected sensor/patroller allocation/reallocation vertex $\nu_{old}$ is changed to another one $(\nu_{new})$, selected randomly:
$M^1_2((V_p, V_s, V_r)) = (V_p^\prime, V_s^\prime, V_r^\prime)$ and $\exists!_z V_z\neq V_z^\prime: z \in \{p, s, r\}$, $V_z = (\nu_1,\ldots, \nu_{old}, \ldots, \nu_k)$, $V_z^{\prime} = (\nu_1, \ldots, \nu_{new}, \ldots, \nu_k)$.

In the case of $M^2_2$, a random element of the signaling probability strategy $\boldsymbol{\Psi^\theta}$ or $\boldsymbol{\Phi^\theta}$ ($\theta \in \{\bar{s}, s^+,s^-\}$) is changed to the \textit{complementary} probability:
$M^2_2(\boldsymbol{\Psi^\theta}) = \boldsymbol{\Psi^{\theta\prime}}$ and $\exists!_{z \in \{1,\ldots,\mathcal{N}\}}$: $\boldsymbol{\Psi^\theta} = [\Psi^\theta_1, \ldots, \Psi^\theta_z, \ldots, \Psi^\theta_{\mathcal{N}}]$, $\boldsymbol{\Psi^{\theta\prime}} = [\Psi^\theta_1, \ldots, 1-\Psi^\theta_z \dots, \Psi^\theta_{\mathcal{N}}]$, with an analogous operation for $\boldsymbol{\Phi^\theta}$. 
By applying a random perturbation, $M_2$ aims to explore new areas of the search space to find potential improvements.

\textit{Coverage improvement} ($M_3: \mathcal{E} \mapsto \mathcal{E}$) is based on information from the evaluation process from the previous generation. In the evaluation procedure, the adversary's optimal strategy is computed (see Evaluation below), which contains a target vertex, $\nu_{adv}$, that the adversary will choose. Thus, a natural idea is to increase the coverage probability of this vertex. This is realized by randomly finding a pure strategy that does not cover $\nu_{adv}$ (i.e., no patroller or sensor is allocated to this vertex), and adding $\nu_{adv}$ either to $V_p$ or $V_s$ in this strategy, in place of another randomly selected vertex $\nu_{rand}$:
$M_3((V_p, V_s, V_r)) = (V_p^\prime, V_s^\prime, V_r)$ and $\exists!_z V_z\neq V_z^\prime: z \in \{p, s\}$, $V_z = (\nu_1, \ldots, \nu_{rand}, \ldots, \nu_k)$, $V_z^{\prime} = (\nu_1, \ldots, \nu_{adv}, \ldots, \nu_k)$ .
Observe that it is not guaranteed that this greedy approach will improve the defender's strategy. In some situations, however, the lack of target coverage is done purposefully and increases the final expected payoff.

If a chromosome is chosen for mutation based on $\mathcal{P}_{m}$, one of the three above described mutation types is randomly chosen and applied to the chromosome (see Figure~\ref{fig:easgs_overview}). This procedure is repeated until a better-fitted chromosome is obtained or $m_{limit}$ trials is reached. After each unsuccessful mutation attempt, the chromosome is reverted to its state before mutation. 
The last mutation attempt is kept even if it leads to a lower-fitted chromosome. In preliminary experiments we observed that, in case none of the mutation attempts returned a better individual, keeping the best mutated individual is not statistically superior to simply using the last resulting individual.

%\subsubsection{Local optimization}
%\label{sec:local-optimization}
%\smallskip
%\noindent
%\textbf{Local optimization.}
%--MOVED to SM --
%{\color{orange}Population initialization, crossover, and mutation are extensively randomized procedures, and some of the resultant encoded strategies may be infeasible or clearly inefficient. In order to fix this issue and improve the results, the following local optimization procedure is executed.  First, for each individual, the algorithm ensures that the encoded solution is feasible, i.e., related vertices from $V_p$ and $V_r$ are connected ($\forall_{{i_k} \in \{1,\ldots,k\}} (\nu_{i_k}^p, \nu_{i_k}^r) \in E$). If this condition is not fulfilled for the $i_k$-th pair, a randomly selected neighbor of vertex $\nu_{i_k}^p$ is assigned as $\nu_{i_k}^r$. 
%Also, there is no need to protect the same vertex with more than one defender's resource (patroller or sensor), so in case of such a multiple assignment, ``spare'' resources are assigned randomly to other unprotected vertices.}

%\subsubsection{Evaluation}
%\label{sec:evaluation}
\subsection{Evaluation}
After population initialization, as well as crossover and mutation, a post-processing step (local optimization) takes place to ensure chromosomes' feasibility. Next, the evaluation procedure calculates the fitness value of each chromosome which is the defender's expected payoff when the mixed strategy encoded in the chromosome is played. Since it is proven in~\cite{conitzer2006computing} that for Stackelberg Games there always exists at least one best adversary's response to the defender's mixed strategy in the form of a pure strategy, it is sufficient to evaluate the defender's strategy against all pure adversary strategies. 
%Due to the relatively small number of these pure strategies, we simply iterate over all of them, compute the game result (players' payoffs), and choose the one that gives the adversary the highest payoff. The defender's payoff against this strategy is assigned as the chromosome fitness value.

%The detection and observational uncertainties introduced in the problem definition % (cf. uncertainty matrix $\Pi$ in Section~\ref{sec:problem-definition}) 
%are considered in the chromosome evaluation process when computing expected payoffs.
%The defender’s strategy evaluation is computed by determining the defender's expected payoff. 
To this end, we first determine the adversary's strategy by iterating over all possible adversary’s pure strategies. Each adversary's pure strategy contains a commitment to allocation (vertex to be attacked) and signaling reaction scheme (response to observed signals - run away or continue the attack).
For each possible pair (a vertex and a signaling reaction scheme), players' payoffs are computed. %Note that circles stating ``adversary is caught,'' ``attack successful,'' and ``attack interrupted'' represent an assignment of the final payoffs. 
Next, we choose the adversary's strategy yielding the best adversary payoff, and determine the defender's payoff for this scenario, which is considered the expected defender's payoff. This value is assigned as the chromosome fitness value.

To compute players' payoffs, we determine the marginal probability $x_\nu^{\theta}$ of the presence of a patroller, a sensor, or no defender resource at the currently attacked vertex based on the list of allocation pure strategies $e$ and their corresponding probabilities $q$.  
Then, for each defender's pure strategy, each of the three possible states ((1)-(3) below) is considered separately and assumed to be the final defender's strategy. The respective defender's payoff is multiplied by the marginal probability. The following three states are considered:
%Each of these states impacts the payoffs as follows: 
(1) In the case of a patroller's presence in the vertex, the defender simply catches the adversary. 
(2) If neither a patroller nor a sensor are located in the attacked vertex, then the resulting payoff depends on the patroller’s reallocation. If a patroller is reallocated to the attacked vertex in the reaction stage, the adversary will be caught.
(3) The last possibility is that a sensor is located in the attacked vertex (and there is no patroller allocated).
In that case, two scenarios must be considered, depending on whether or not the sensor detects the adversary. The probability of each scenario is defined by $\gamma$, the detection uncertainty. 
The probability of sending signal $\sigma_0$, given by $\boldsymbol{\Psi^\theta}$ and $\boldsymbol{\Phi^\theta}$, depends on several factors: the adversary's detection, the signaling strategy, and the reallocation state (whether there is a patroller in the neighborhood that will visit this vertex in the reaction stage). All the above information leads to computing the probability of a sensor’s signaling strategy\footnote{ chromosome's identifier ($j$) is omitted for clarity in equations}:
$$\mathcal{P}_{\sigma_1}(\nu) = \gamma \sum_\theta  (x_\nu^\theta \Phi_\nu^\theta) + (1-\gamma) \sum_\theta (x_\nu^\theta \Psi_\nu^\theta)$$

$$\mathcal{P}_{\sigma_0}(\nu) = \gamma \sum_\theta x_\nu^\theta (1-\Phi_\nu^\theta) + (1-\gamma) \sum_\theta x_\nu^\theta (1-\Psi_\nu^\theta)$$

where 
$\mathcal{P}_{\sigma_1}(\nu)$ and $\mathcal{P}_{\sigma_0}(\nu)$ are probabilities of sending signals $\sigma_1$ and $\sigma_0$ in vertex $\nu$, respectively,\\
$\theta \in \{\bar{s}$, $s^+$, $s^-$\} are allocation states,\\
$x_\nu^\theta$ represents the marginal probability that vertex $\nu$ is in the allocation state $\theta$.\\
$x_\nu^\theta$ is computed based on allocation strategies $(e_i,q_i)$, for example: $x_\nu^{s^+} = \sum q_i : \nu \in V_r \subset e_i$.

The next step is to determine the adversary's reaction.
Based on the signaling strategy and probabilities of how certain signals can be observed by the adversary ($\Pi$), the adversary's reaction is computed which, together with the defender's strategy, allows for computing the final payoffs.

The detailed diagram of the evaluation procedure is presented in supplementary material.

%To compute the payoff, marginal probabilities of the 3 possible states are calculated (based on the allocation pure strategies $e$ and their corresponding probabilities $q$): (1) presence of a patroller, (2) presence of a sensor, or (3) no defender's resources in the attacked vertex. In cases (1) and (3), the game ends with the adversary being caught or with a successful attack, resp. and related payoffs are calculated. For state (2), a probability of sending a signal is computed based on the sensor’s detection uncertainty $\gamma$ and signaling strategy ($\boldsymbol{\Phi^\theta_j}$ and $\boldsymbol{\Psi^\theta_j}$), encoded in the chromosome. Then, based on the adversary's reaction, the final payoffs are calculated.
%{\color{red}Further details are in the supplementary material.}

%\subsubsection{Selection}
%\label{sec:selection}
\subsection{Selection}
In order to construct a population for the next generation, the selection procedure is executed on a pool of all parent individuals (the current population), and the results of crossover and mutation operators (offspring individuals). In the beginning of the selection process, a fixed number, $n_{e}$ of \emph{elite} individuals from this pool are automatically promoted to the new population. The elite individuals are %The elite pool consists of $n_{e}$ 
top-ranked chromosomes with the highest fitness values. %, transferred unconditionally to preserve the best solutions found so far.  

Next, the following iterative procedure is executed until the next generation population is filled with $n_{pop}$ individuals. In each iteration, two chromosomes are randomly drawn with return from the pool. Then, from this pair, the one with a higher fitness value is promoted to the next generation with probability $\mathcal{P}_{sp}$ (selection pressure). Otherwise, the lower-fitted one is promoted.

%\subsubsection{Stop condition}
%\label{sec:stop-condition}
\subsection{Stop condition}
Multiple applications of the proposed operators can, in principle, lead to any arbitrary solution. Repeated crossover can create a mixed strategy with an arbitrary number of pure strategies, $M_1$ can set any probability distribution of these strategies, and $M_2$ applied multiple times is able to transform each arbitrary pure strategy to any other. Hence, any mixed strategy can be potentially achieved through repeated application of these operators, independently of the initial population selection. 
%Consequently, given enough time the algorithm can converge to the optimal solution. 
However, it is difficult to say anything about theoretical time requirements of the method. Providing theoretical convergence guarantees for EAs is generally a challenging task and formal results in this area occur rarely in the literature.
Thus, the whole routine (crossover, mutation, local optimization, evaluation, and selection) is repeated until some fixed number of generations $n_{gen}$ is reached. 

%However, we observed that algorithm sometimes tends to stuck in some areas and the best solution is not %changed. Thus, in order to mitigate this issue, the following procedure is implemented. 
Furthermore, if for $n_{ref}$ (refresh) generations a payoff of the top-ranked individual does not change, some fresh chromosomes are added to the population, i.e., we randomly select half of the population, which is replaced by the same number of new individuals with randomly assigned pure strategies, generated in the same way as in the initial population. The best individual is preserved and cannot be deleted. This approach avoids getting stuck in sub-optimal solutions and shifts the population to new potentially better areas. %{\color{red}Further discussion about the role of this \textit{refreshing} procedure and performance visualisation can be found in the supplementary material.}

\section{Experimental setup}
\label{sec:setup}
Access to the data from the real-world scenario is strictly secured, e.g. to protect the locations of wild animals and rangers resources, so we randomly generated 342 artificial games 
%with settings similar to those observed in practice - for instance, usually there are not too many sensors/patrollers because they are expensive.}
%{\color{orange}To protect the locations of wildlife and patroller resources, we created a benchmark of 342 random games 
with various utilities, graph architectures, numbers of vertices, patrollers, and sensors.
All generated games are publicly available on github.com/easgs/benchmark\_games.

EASGS was compared with the exact approach, %Signaling Exponential Linear Programming 
SELP, and the three heuristic methods: % - Signaling Branch and Price (
SBP, %Signaling Branch and Price with Warm-up (
SBP+W and m-CombSGPO, summarized in Related Work. 
%All three baselines were proposed in \cite{bondi2020signal}. 
Results for non-deterministic algorithms (including EASGS) were obtained in $30$ independent runs per game instance. Tests were performed on a cluster running CentOS Linux 7 (Core) with Intel(R) Xeon(R) CPU E5-2683 v4 @ 2.1 GHz with 
%up to 
128 GB RAM and 4 cores.
EASGS source code can be found on github.com/easgs/source\_code.

%\subsubsection{Benchmark games}
\subsection{Benchmark games}
%--MOVED TO SM -- {\color{orange}To generate our benchmark games, we implemented the following procedure.  %in~\cite{bondi2020signal}. %, where Signaling Security Games were introduced. 
%Utilities assigned to vertices were generated randomly with
%with a maximum absolute value equal to $1090$,
%the defender's utilities for catching the adversary being smaller than the adversary's utilities for a successful attack. Their magnitudes are further justified in~\cite{bondi2020signal}, e.g. animals are worth more for ecotourism than for sale on the black market. The number of patrollers and sensors depended on the number of graph vertices ($\mathcal{N}$) and equaled $k=\sqrt{\frac{\mathcal{N}}{2}}$ and $l=\frac{2}{3}\mathcal{N}-k$, resp. Detection and observational uncertainty, parameterized by $\gamma$ and $\kappa$, were set randomly and uniformly: $\gamma, \kappa \in [0, 1]$. Details can be found in the supplementary material.}
%
Let $\mathcal{N}$ and $a_d$ denote the graph size and the average vertex degree. Games were generated within four groups: \emph{sparse} (50 games), with $a_d = 2$, \emph{moderate} (50 games) ($a_d = \mathcal{N}/2$), \emph{dense} (50 games) ($a_d = \mathcal{N}-2$) and \emph{locally-dense} (192 games), a set of connected cliques. 

Sparse, moderate, and dense game graphs were random Watts-Strogatz graphs with $\mathcal{N}=10, 20, \ldots, 100$. 
%For each $\mathcal{N}$, $5$ different games (in terms of graph structure and utilities) were generated. Thus, }in total $50$ games were created for each graph type.
In \emph{locally-dense} graphs, the number of cliques and their size varied from 3 to 10. 
%Additionally, 3 types of connections between cliques were considered.
%} 
%Overall, $192$ different 
\emph{locally-dense} games 
%were generated. This type of a graph is 
are inspired by certain real-life scenarios, in which there exist areas (cliques) which patrollers can quickly explore, and these are connected via larger links, e.g., roads, which speed up travel. In \cite{xu2020stay}, patrollers travel to areas via motorbikes and then patrol on foot.
The graphs are further described in supplementary material. Figure~\ref{fig:sample-games} presents four examples.  % for a detailed description of their generation procedure and more visualizations. 
%{\color{orange}\emph{Locally-dense} graphs are another important contribution of this paper since previous works (e.g. \cite{bondi2020signal}) did not account for the possibility of different travel times between areas.}

%In total, EASGS was evaluated on $342$ games with various utilities, graph architectures, numbers of targets (vertices), patrollers, and sensors.

\begin{figure}[ht]
\centering
  \subfloat[sparse]{
    \includegraphics[width=.47\columnwidth]{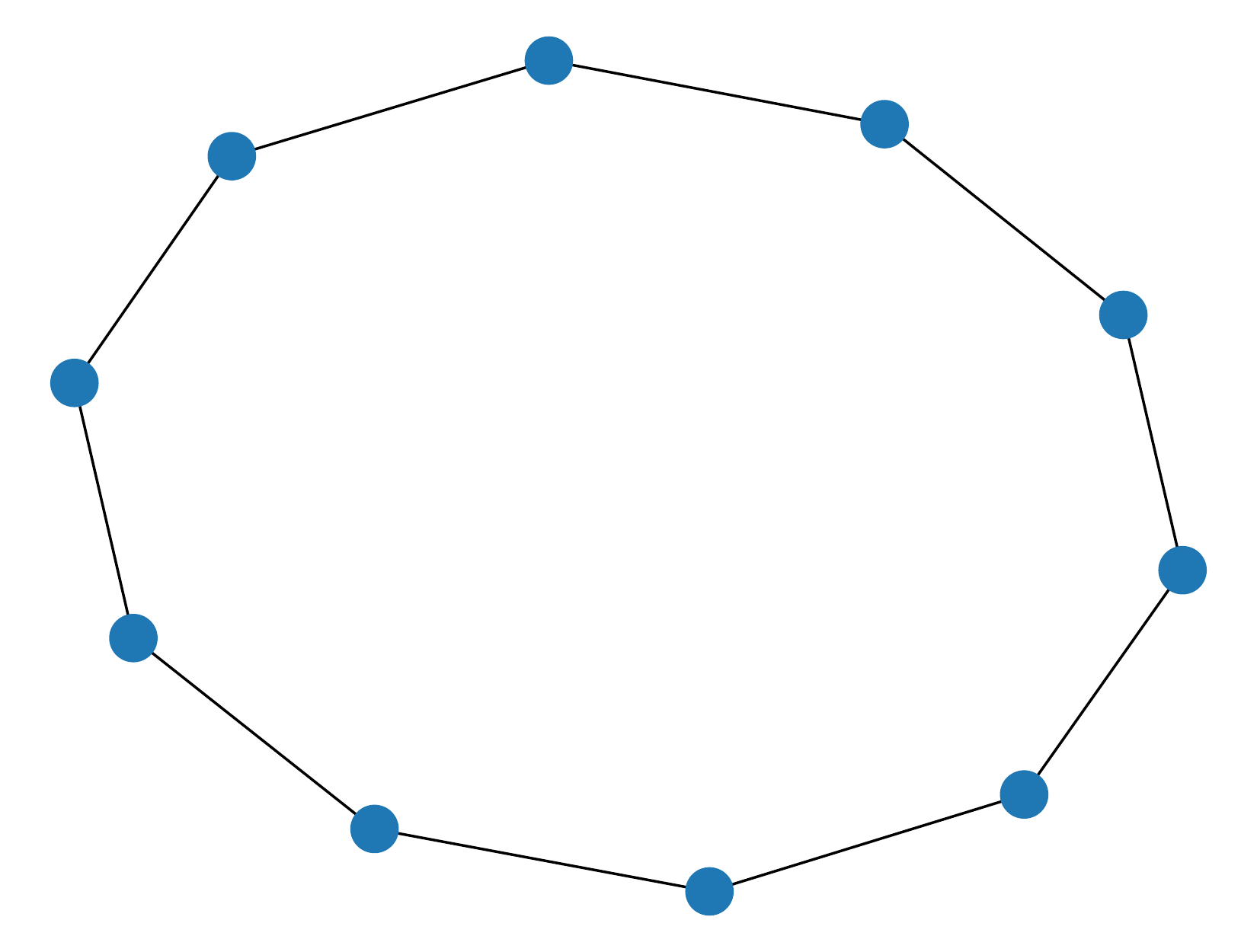}%
  }
  \subfloat[moderate]{
    \includegraphics[width=.47\columnwidth]{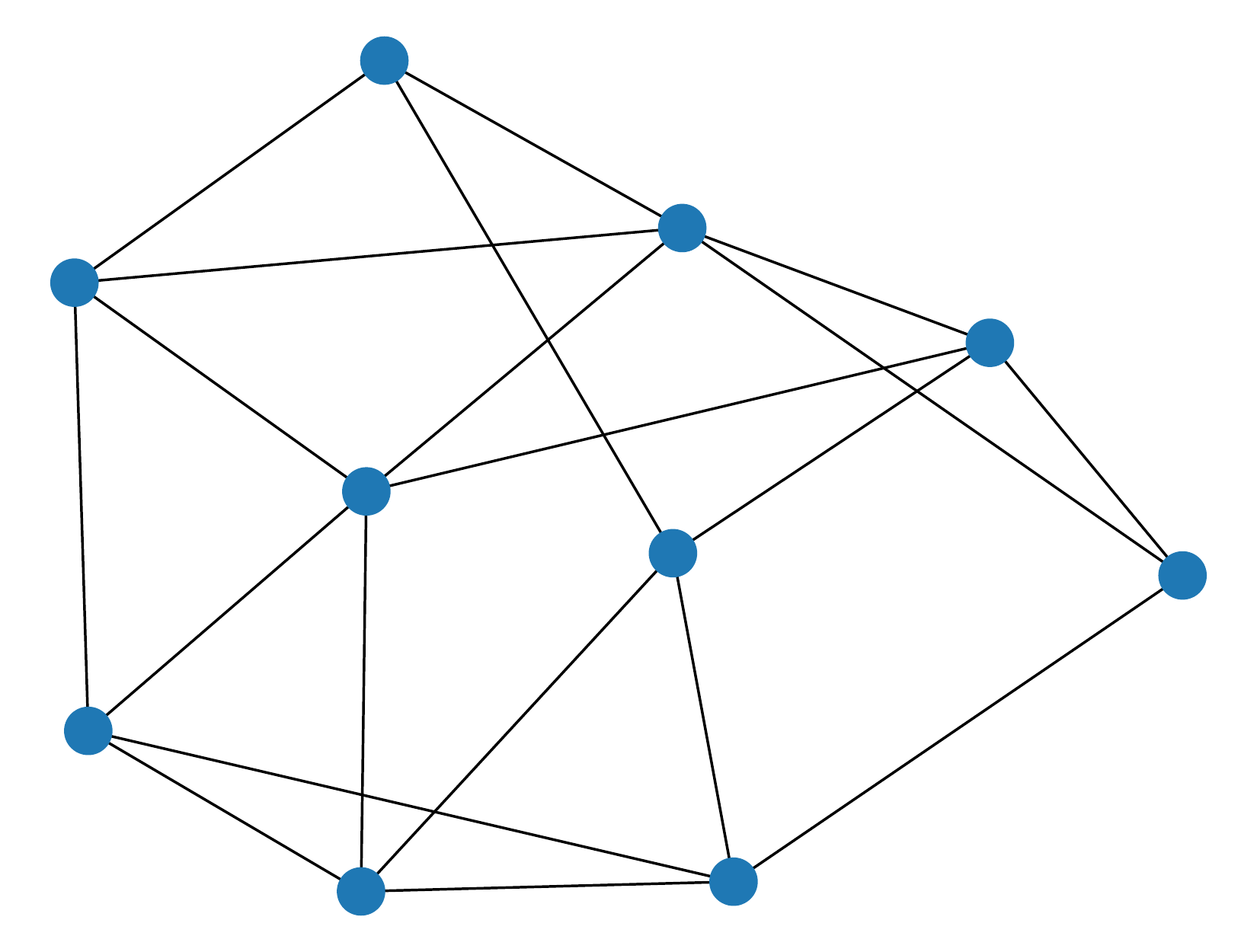}%
  }
  \quad
  \subfloat[dense]{
    \includegraphics[width=.47\columnwidth]{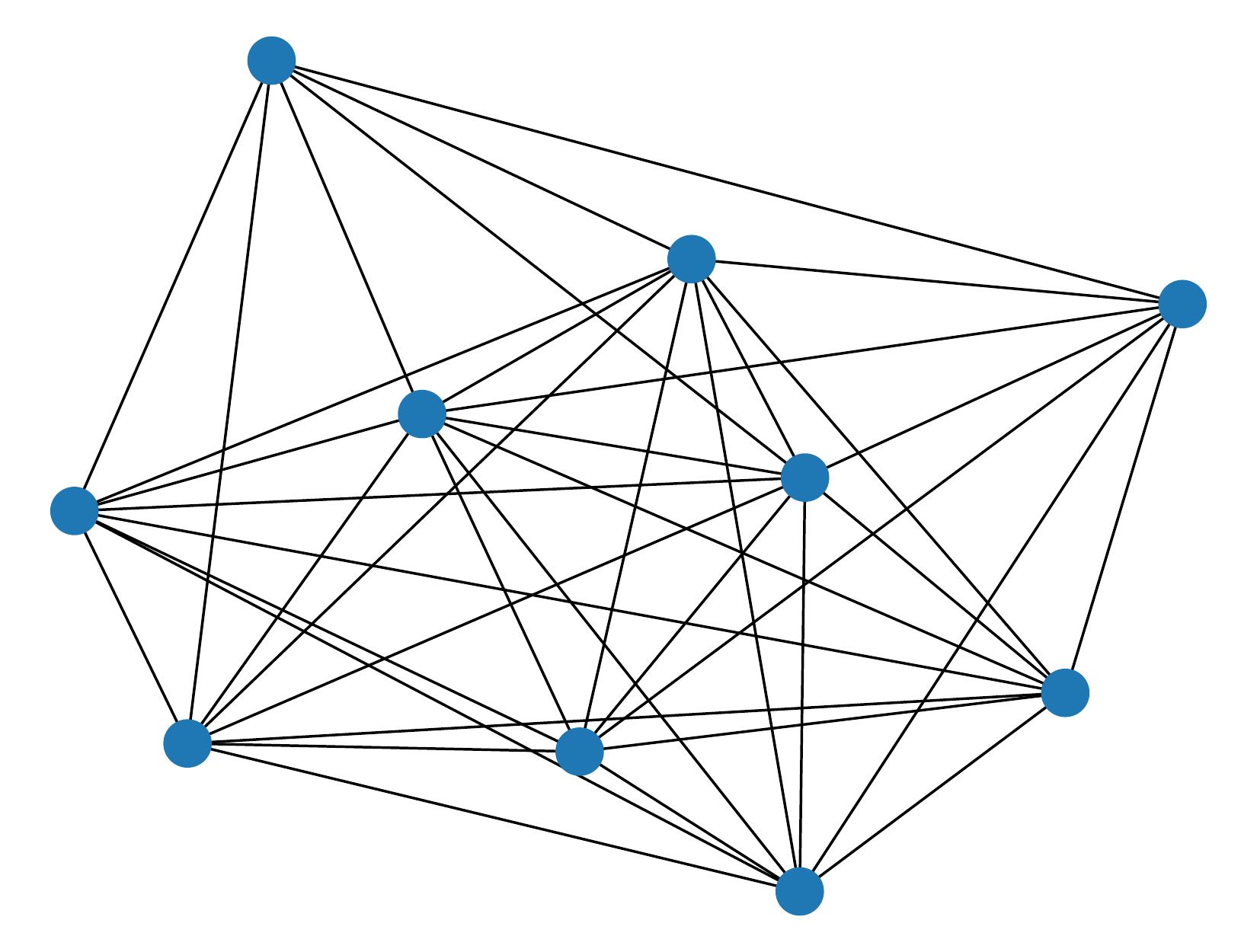}%
  }
  \subfloat[locally dense]{
    \includegraphics[width=.47\columnwidth]{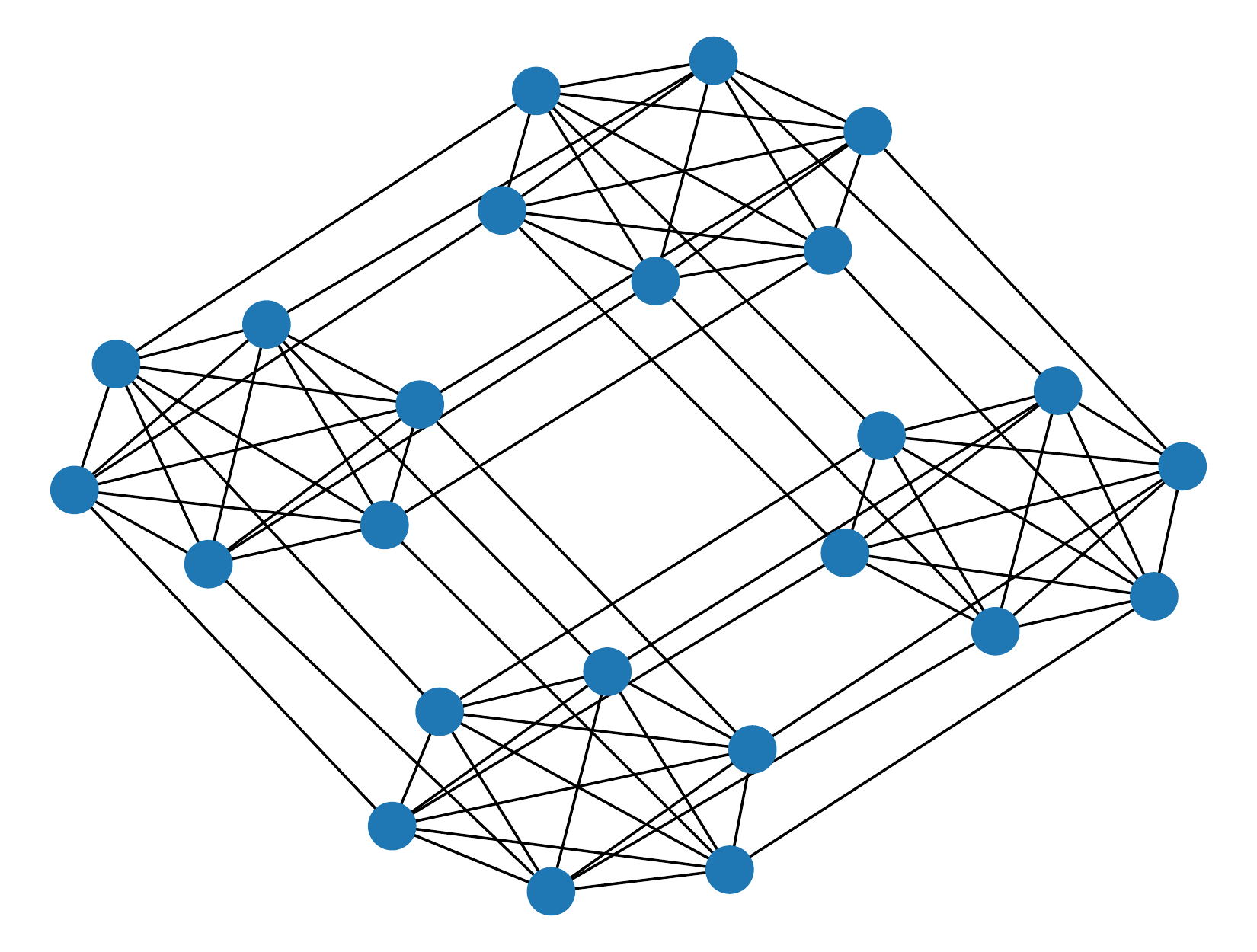}%
  }
  \caption{Sample benchmark game graphs.}%
  \label{fig:sample-games}
\end{figure}

%\subsubsection{EASGS parametrization}
\subsection{EASGS parameterization}
EASGS parameters were tuned on a set of $12$ games with $20$ vertices ($3$ games of each type: \textit{sparse}, \textit{moderate}, \textit{dense} and \textit{locally-dense}), which were separated from the 342 EASGS benchmark graphs and not used during the method evaluation.
Based on $5000$ runs, the following parameter values were finally chosen: population size $n_{pop} = 200$, crossover probability $\mathcal{P}_{c} = 0.5$, mutation probability $\mathcal{P}_{m} = 0.8$, mutation repetition limit $m_{limit} = 10$, number of elite chromosomes $n_{e} = 2$, selection pressure $\mathcal{P}_{sp} = 0.8$, generations limit $n_{gen} = 2000$, number of generations between refreshes $n_{ref} = 300$. 
%{\color{orange}Detailed results of the EASGS parameterization process and more comprehensive discussion can be found in the supplementary material.}

\section{Results and Discussion}
\label{sec:results}
The results are averaged over 30 independent runs and presented in 3 perspectives: payoffs, time scalability, and memory requirements.

\begin{figure*}[ht!]
\centering
  \subfloat{
    \includegraphics[width=0.79\hsize]{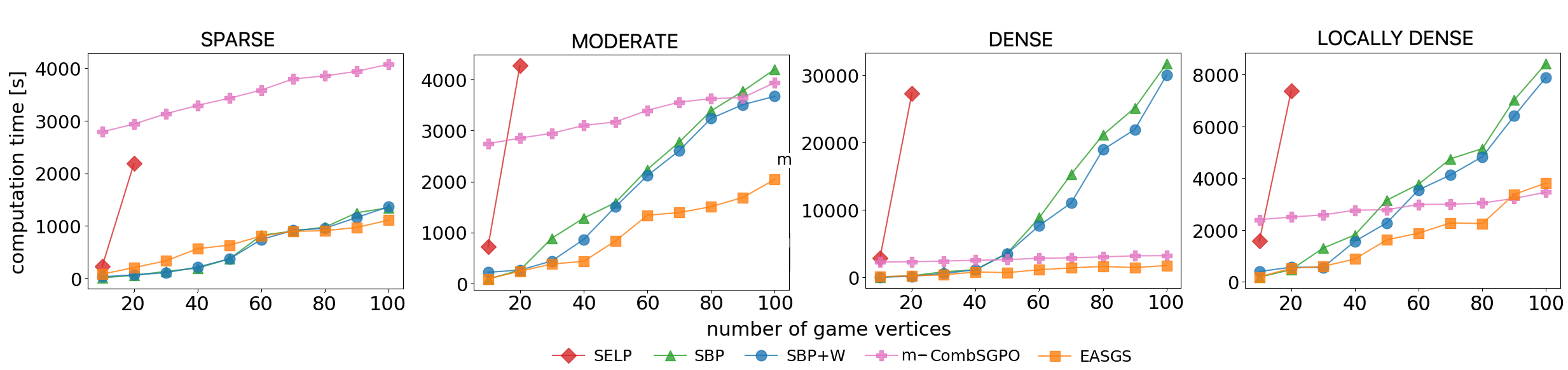}%
    \label{fig:time_scalability}
  }
  \subfloat{
    \includegraphics[width=0.2\hsize]{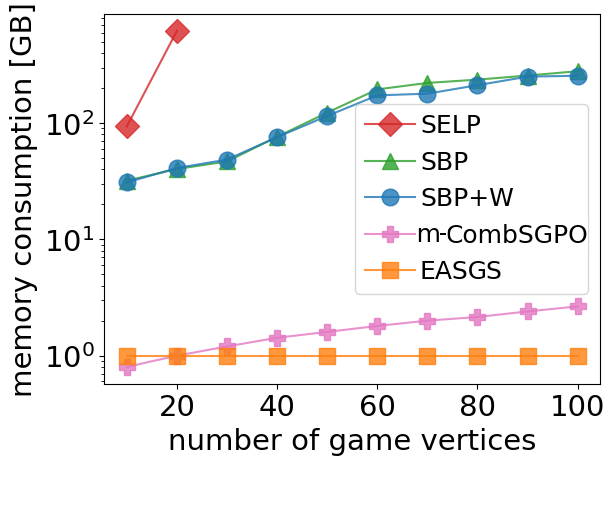}%
    \label{fig:memory_consumprion}
  }
  \caption{Time (left) and memory (right) scalability averaged over 5 games (m-CombSGPO had to be trained separately for each game).}
  \label{fig:time_and_memory}
\end{figure*}

%\subsubsection{Payoffs}
\subsection{Payoffs}
SELP managed to calculate the payoffs only for games with up to 20 vertices. For larger games, its memory consumption exceeded the limit of 128 GB. Thus, we were able to compare obtained results with the optimal solution only for 30 game instances - 6 for each of \textit{sparse}, \textit{moderate} and \textit{dense} type, and 12 \textit{locally-dense} ones. Results for these games are presented in Table~\ref{tab:results_optimal}.

\begin{table}[ht]
\small
\setlength\tabcolsep{4px}
\resizebox{1.0\columnwidth}{!}{
  \begin{tabular}{c|c|c|c}
 & SBP & SBP+W & EASGS \\\hline
\textit{sparse} & \; \textbf{0.00 $\mid$ 0.00 $\mid$ 6/6} & \textbf{0.00 $\mid$ 0.00 $\mid$ 6/6} & 0.90 $\mid$ 0.01 $\mid$ 1/6 \\
\textit{moderate} & 15.77 $\mid$ 0.17 $\mid$ 1/6 & 7.92 $\mid$ 0.09 $\mid$ \textbf{2/6} & \textbf{1.51 $\mid$ 0.02} $\mid$ 1/6 \\
\textit{dense} & 18.93 $\mid$ 0.26 $\mid$ \textbf{0/6} & 9.25 $\mid$ 0.09 $\mid$ \textbf{0/6} & \textbf{1.98 $\mid$ 0.03 $\mid$ 0/6} \\
\textit{locally-dense} & \; 13.45 $\mid$ 0.28 $\mid$ 1/12 & \; 3.16 $\mid$ 0.05 $\mid$ \textbf{4/12} & \; \textbf{1.42 $\mid$ 0.02} $\mid$ 3/12 \\
  \end{tabular}
}
\caption{In order to make a thorough comparison, we compare the best EASGS, SBP, SBP+W solutions with the optimal ones in terms of both the average difference (left columns) and the ratio between the defender's payoffs
%returned by a given method and the exact algorithm 
(middle columns). Right columns present fractions of games for which optimal solutions were found. Each method is assumed to have reached the optimal result if the returned payoff difference smaller than $0.01$. The best results are bolded.}
\label{tab:results_optimal}
\end{table}

SBP and SBP+W reached optimal solutions in all \textit{sparse} test games. However, for denser graphs, their accuracy degrades significantly, and overall, the leading method is EASGS. At the same time, EASGS reaches the exact optimal result less frequently (5 out of 30 games). In other words, EASGS is able to find areas with high quality solutions, but may have some difficulties with precise exploitation of those areas. 
%Further exploration of this phenomenon is left for our future work. 
%{\color{orange}We hypothesize that due to SGS specificity, a gradual decrease of the exploration to exploitation balance as the EASGS progresses could be beneficial.} 
m-CombSGPO (which is based on reinforcement learning) is not a suitable method for finding exact solutions. It has not reached the optimal solution for any game instance and therefore is omitted in Table~\ref{tab:results_optimal}.
%This is a characteristic behavior of those %evolutionary algorithms which are more directed %towards exploration than %exploitation~\cite{vcrepinvsek2013exploration}. %EASGS can find areas of the strategy space with %good quality solutions, but may have some %difficulties with precise exploitation of those %areas. 
%This proposed method's feature follows a huge search space (e.g. more than $10^150$ possible pure %strategies for games with $100$ vertices).

\begin{table}[ht]
\small
\setlength\tabcolsep{4px}
\resizebox{1.0\columnwidth}{!}{
  \begin{tabular}{c|c|c|c|c}
 & SBP & SBP+W & m-CombSGPO & EASGS \\\hline
\textit{sparse} & -86.68 (84\%) & \textbf{-86.01 (92\%)} & -419.86 (0\%) & -91.32 (6\%) \\
\textit{moderate} & -75.01 (2\%) & -72.75 (36\%) & -255.73 (0\%) & \textbf{-69.92 (62\%)} \\
\textit{dense} & -58.72 (2\%) & -57.98 (34\%) & -149.14 (0\%) & \textbf{-51.47 (64\%)} \\
\textit{locally-dense} & -60.68 (4\%) & -57.80 (26\%) & -340.65 (0\%) & \textbf{-54.36 (70\%)} \\
  \end{tabular}
}
\caption{Comparison of the average defender's payoff across all benchmark games. In brackets are the percentages of games for which a given method obtained the winning result. Note that multiple methods could achieve the same winning result, so the scores may not sum to 100\%. The best results are bolded.}
\label{tab:results_all}
\end{table}

Table~\ref{tab:results_all} shows a comparison of heuristic methods based on all $342$ test games.
The results confirm that EASGS works better for all types of games except for \textit{sparse} ones, whereas both SBP and SBP+W degrade on denser graphs. The reason may lie in the increasing branching factor of denser games, as each additional connection between vertices opens the possibility for patroller reallocation and enlarges the game strategy tree. In all cases m-CombSGPO reaches the lowest payoffs.
Overall, for 200 out of 342 test game instances, EASGS yielded the best result. For each type of benchmark game, the advantage of the winning method (SBP+W for \textit{sparse} games and EASGS for all other types) is statistically significant according to a 1-tailed paired t-test with a significance level equal to $0.05$, and with a normal distribution of data checked by a Shapiro-Wilk test. However, when all $342$ games of all types are considered together, it is not statistically significant, with the significance level equal to $0.08$.
EASGS is also a stable method. The average standard deviation (30 runs of each game) equals 0.86 (about 1.2\% of the average absolute payoff - cf. Table~\ref{tab:results_all}) with the maximal value of 1.63.
%which shows the proposed method ability to reproduce good results.}

%\subsubsection{Time scalability}
\subsection{Time scalability}
%\begin{figure}[ht]
%	\begin{center}
%    \includegraphics[width=1.0\columnwidth]{images/time-scalability.png}
%    \caption{Time scalability with respect to the number of vertices.}
%    \label{fig:time_scalability}
%	\end{center}
%\end{figure}
%
Another critical aspect of the proposed algorithm is time scalability. Figure~\ref{fig:time_and_memory} compares time efficiency of the considered methods. Except for the smallest game graphs and moderate-size \textit{sparse} ones (where SBP+W performs slightly better), EASGS outperforms all competitive methods, sometimes by a large margin (e.g. on \textit{dense} graphs). The explanation of this phenomenon is related to the way the EASGS operations are implemented. Action complexity in EASGS is independent of the underlying graph density: mutation assigns vertices to patrollers/sensors randomly, and the topology of connections between vertices (and the number of them) do not influence those choices. %For m-CombSGPO, the times reported are the training times for each graph, as we had to train m-CombSGPO separately for each graph. 
%{\color{olive}Each point in time graph was the average time for 5 games with the number of nodes specified on the x axis. m-CombSGPO had to be trained separately for each of those 5 games.}%Topology of connections are relevant only in the local optimization phase, but still the vertex for patroller reallocation it selected randomly from the list of neighbors and complexity of this operation does not depend on the number of connections (neighbors).

%\subsubsection{Memory consumption}
\subsection{Memory consumption}
Although memory consumption is typically not a key factor in recent years, in some cases (especially for methods based on traditional optimization techniques), memory utilization can hinder real-life applications. SELP, for example, %can compute only games with a small number of vertices because of exponential time scalability. It may be possible to calculate the solution for larger games in a reasonable time, but even for fairly small games with 20 vertices, SELP 
consumes more than 200 GB of memory for small games with 20 vertices. At the other extreme, EASGS memory utilization is almost constant. The only data stored in EASGS memory is the game definition and the EA population. %{\color{violet}Each individual in the population is represented by a set of real vectors.} 
This is reflected in 
Figure~\ref{fig:time_and_memory}, which presents memory utilization for all tested methods. 
%
%\begin{figure}[ht]
%	\begin{center}
%    \includegraphics[width=0.7\columnwidth]{images/memory_consumption.pdf}
%    \caption{Comparison of EASGS memory consumption vs state-of-the-art methods.}
%    \label{fig:memory_consumprion}
%	\end{center}
%\end{figure}
%
%EASGS memory consumption is negligible compared to other methods and 
Even for 100-vertex games, EASGS memory does not exceed 150 MB. 
%Thus, EASGS is very efficient in terms of memory consumption and it is a clear advantage over other %state-of-the-art MILP solutions.
%Efficient memory utilization is a clear advantage of EASGS in comparison with traditional optimization approaches.

\subsection{Qualitative analysis}
%During detailed qualitative analysis of the individual outcome EASGS defender strategies, we observed some common characteristics. 
The majority of EASGS resultant mixed strategies contained between 4 to 8 pure strategies (with the mean of 5.3). There wasn't a single resultant strategy with an uncovered vertex, i.e. for each vertex there was at least one pure strategy with a sensor or patroller assigned to that vertex, meaning the target coverage generated by EASGS was complete. Moreover, the patrollers were generally assigned to higher degree vertices, 
%(i.e., those with more connections), 
which allowed patrollers to respond to more targets (checked by sensors) during the reallocation stage. Targets with higher defender penalties were generally covered with higher probability. EASGS also considered adversary utilities by protecting targets of high value for the adversary, as well as graph architecture by assigning the vast majority of the sensors to vertices that are connected to targets with assigned patrollers. %{\color{red}An example visualization of a resulting strategy and further discussion are presented in the supplementary material.}% Overall, all analysed strategies look reasonably and it is very difficult to indicate any single point of improvement (e.g. by changing some pure strategy probability or reallocating defender resources).

%\subsubsection{Ablation study}
%\label{sec:ablation}
%--MOVED TO SM-- \smallskip
%\noindent
%\textbf{Ablation study.}
%{\color{olive}EASGS implements several novel components compared to standard EA realization, including three types of mutation, mutation repetition, the removal of strategies in crossover, specific population \textit{refreshing}, and local coverage optimization.
%population \textit{refreshing}.
%and it may be not obvious what are their roles, how they impact the final result. 
%In order to better understand the role of particular enhancements and their actual impact on the overall EASGS performance, several ablation experiments were performed. Details can be found in the supplementary material, but generally removing of any of the EASGS components worsens the results.}

\section{Conclusion}
\label{sec:conclusions}
The paper introduces a novel approach to solving Security Games with Signaling based on the Evolutionary Computation paradigm. The method maintains a population of potential solutions (mixed strategies) and modifies them using specially-designed operators which address domain characteristics.
%EASGS robustness was proven by extensive experiments with games with a wide range of properties. 
%
A comprehensive experimental evaluation confirmed that EASGS provides better payoffs than state-of-the-art methods, especially for \textit{locally-dense} graphs that are inspired by real-world settings. Additionally, EASGS largely outperforms competitive methods in terms of time scalability. %, except for a subset of the smallest games and moderate-size \textit{sparse} ones.
%Since EASGS implements other non-MILP approach to SG its memory utilization is significantly less than %state-of-the-art methods and does not limit the algorithm’s usage to small games. 
%
Thanks to nearly constant memory scalability and % \emph{anytime} characteristics (i.e., 
the ability to return a valid solution at any time during the execution process, EASGS can be employed to larger games that are beyond the capacity of the state-of-the-art methods.
Being well-suited for field computing requirements, EASGS takes us a step closer towards deploying real-world applications on a larger scale to protect valuable natural resources and biodiversity. As future work, we plan to test how well EASGS performs in the field by integrating it with more advanced sensors based on real-world object detection from images.

\section*{Acknowledgements}
%Adam {\.Z}ychowski and Jacek Ma{\'n}dziuk were supported by POB Research Centre Cybersecurity and Data Science of Warsaw University of Technology within the Excellence Initiative Program - Research University (ID-UB).
%
The project was funded by POB Research Centre Cybersecurity and Data Science of Warsaw University of Technology within the Excellence Initiative Program - Research University (ID-UB).

\vspace{0.3em}
\noindent
Research was sponsored by the Army Research Office and was accomplished under Grant Number: W911NF-17-1-0370. The views and conclusions contained in this document are those of the authors and should not be interpreted as representing the official policies, either expressed or implied, of Army Research Office or the U.S. Government. The U.S. Government is authorized to reproduce and distribute reprints for Government purposes notwithstanding any copyright notation herein.

\bibliographystyle{named}
\bibliography{EASGS}

\clearpage
\appendix

\begin{center}
\LARGE
\textbf{--- Supplementary material ---}
\end{center}
\vspace{0.3cm}

\section{EASGS evaluation procedure}
%The game scenario follows the Security Games with Signaling (SGS)~\cite{bondi2020signal} and is summarized in Figure~\ref{fig:game-timing}.

\begin{figure*}
\centering
\includegraphics[width=.9\textwidth]{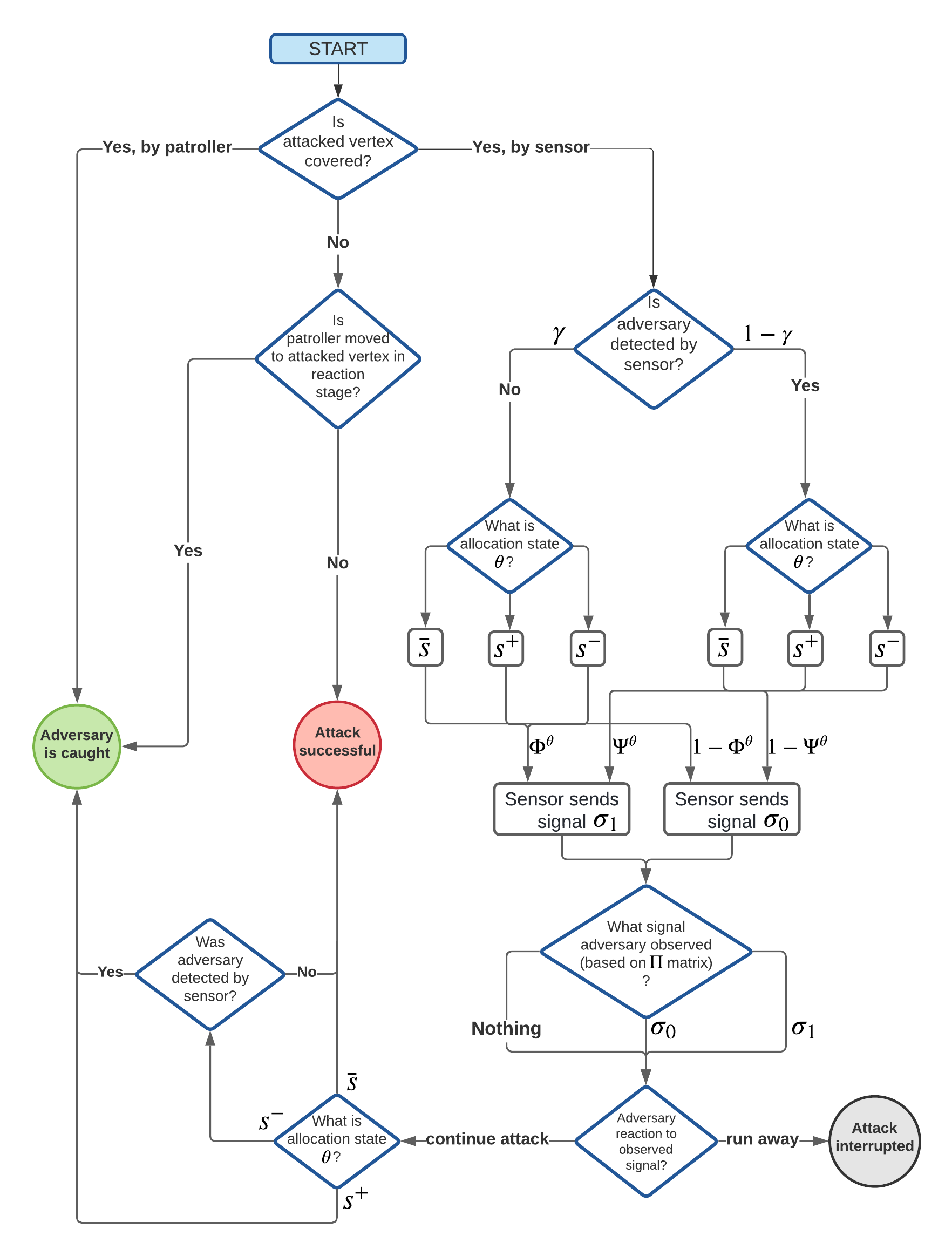}%
\caption{EASGS evaluation workflow.}
\label{fig:EASGS_workflow}
\end{figure*}

In SGS two sources of uncertainty are defined: \textit{detection uncertainty} and \textit{observational uncertainty} (please see the main paper for further description).
Both types of those uncertainties are considered in the EASGS chromosomes' evaluation procedure.
The flowchart presented in Figure \ref{fig:EASGS_workflow} shows consecutive EASGS evaluation steps. Circles stating ``adversary is caught,'' ``attack successful,'' and ``attack interrupted'' represent an assignment of the final payoffs.

\section{EASGS local optimization} Population initialization, crossover and mutation are extensively randomized procedures and some of the resultant encoded strategies may be infeasible or clearly inefficient. In order to fix this issue and improve the results, the following local optimization procedure is executed.  First, for each individual, the algorithm ensures that the encoded solution is feasible, i.e., the related vertices from $V_p$ and $V_r$ are connected ($\forall_{{i_k} \in \{1,\ldots,k\}} (\nu_{i_k}^p, \nu_{i_k}^r) \in E$). If this condition is not fulfilled for the $i_k$-th pair, a randomly selected neighbor of vertex $\nu_{i_k}^p$ is assigned as $\nu_{i_k}^r$. 
Also, there is no need to protect the same vertex with more than one defender's resource (patroller or sensor), so in case of such a multiple assignment, ``spare'' resources are assigned randomly to other unprotected vertices.

% \clearpage 
\section{Benchmark games}

Access to the data from the real-world scenario is strictly secured, e.g. to protect the locations of wild animals and rangers resources. Hence, we randomly generated artificial games with settings similar to those observed in practice. For instance, usually there are not too many sensors/patrollers because they are expensive. 
The created benchmark set includes 342 games with variable utilities.

The number of patrollers and sensors depended on the number of graph vertices ($\mathcal{N}$) and equaled $k=\sqrt{\frac{\mathcal{N}}{2}}$ and $l=\frac{2}{3}\mathcal{N}-k$, resp. Detection and observational uncertainty, parameterized by $\gamma$ and $\kappa$, were uniformly sampled: $\gamma, \kappa \in [0, 1]$.

%We implemented the following generation procedure.
Utilities assigned to vertices were generated randomly with the defender's utilities for catching the adversary being smaller than the adversary's utilities for a successful attack, similar to~\cite{xu2015exploring}. Their magnitudes are further justified in~\cite{bondi2020signal} where they took elephants as an example animal and used the approximate price of ivory, the approximate daily monetary benefit of ecotourism, and an elephant poaching fine to assign payoffs to nodes. The defender payoffs are related to the ecotourism benefits of elephants. The attacker payoffs are related to the price of ivory (successful attack), and the fines associated with a covered target (attack failed). In general, animals are worth more for ecotourism than for sale on the black market.

Benchmark games are divided into 4 categories based on the underlying graph structure: \textit{sparse}, \textit{moderate}, \textit{dense}, and \textit{locally-dense}.
Each of \textit{sparse}, \textit{moderate} and \textit{dense} subsets contains 50 game instances. Graphs were generated randomly according to the Watts-Strogatz model with the number of vertices $|V|= 10, 20, \ldots, 100$. For each graph category and each number of vertices, 5 game graphs (games) were created. 
Thus, the benchmark set contained 50 games per each of the above 3 categories. 
%Sample graphs with 10 vertices are presented in Figure~\ref{fig:sample-games}.

%\begin{figure}[ht]
%\centering
%  \subfloat[sparse]{
%    \includegraphics[width=.47\columnwidth]{images-supplementary/game-sparse.pdf}%
%    \label{fig:game-sparse}
%  }
%  \quad
%  \subfloat[moderate]{
%    \includegraphics[width=.47\columnwidth]{images-supplementary/game-moderate.pdf}%
%    \label{fig:game-moderate}
%  }
%  \quad
%  \subfloat[dense]{
%    \includegraphics[width=.47\columnwidth]{images-supplementary/game-dense.pdf}%
%    \label{fig:game-dense}
%  }
%  \caption{Sample benchmark game graphs.}%
%  \label{fig:sample-games}
%\end{figure}

\begin{figure}[ht]
\centering
  \subfloat[game 03\_06\_1]{
    \includegraphics[width=.47\columnwidth]{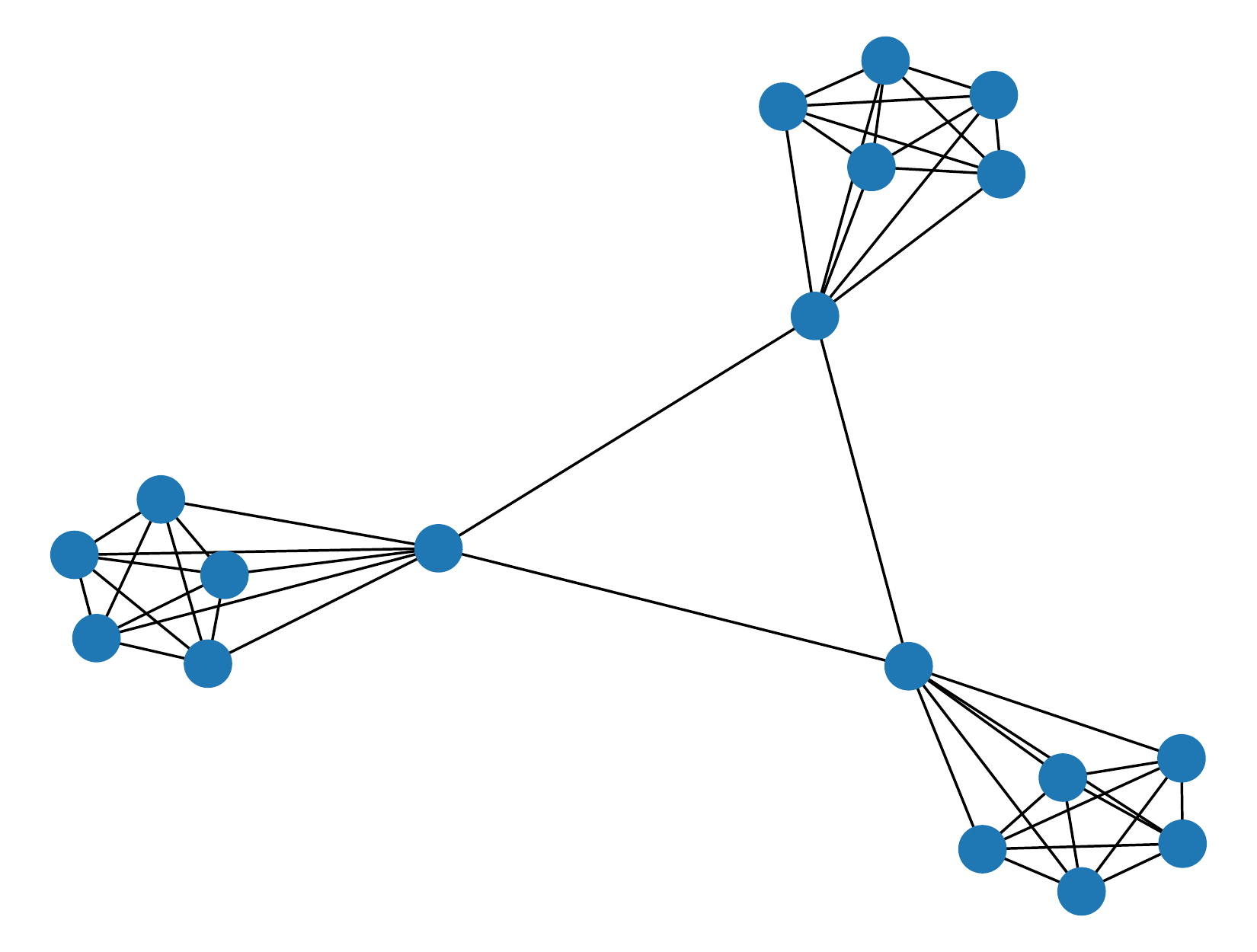}%
  }
  \subfloat[game 04\_05\_1]{
    \includegraphics[width=.47\columnwidth]{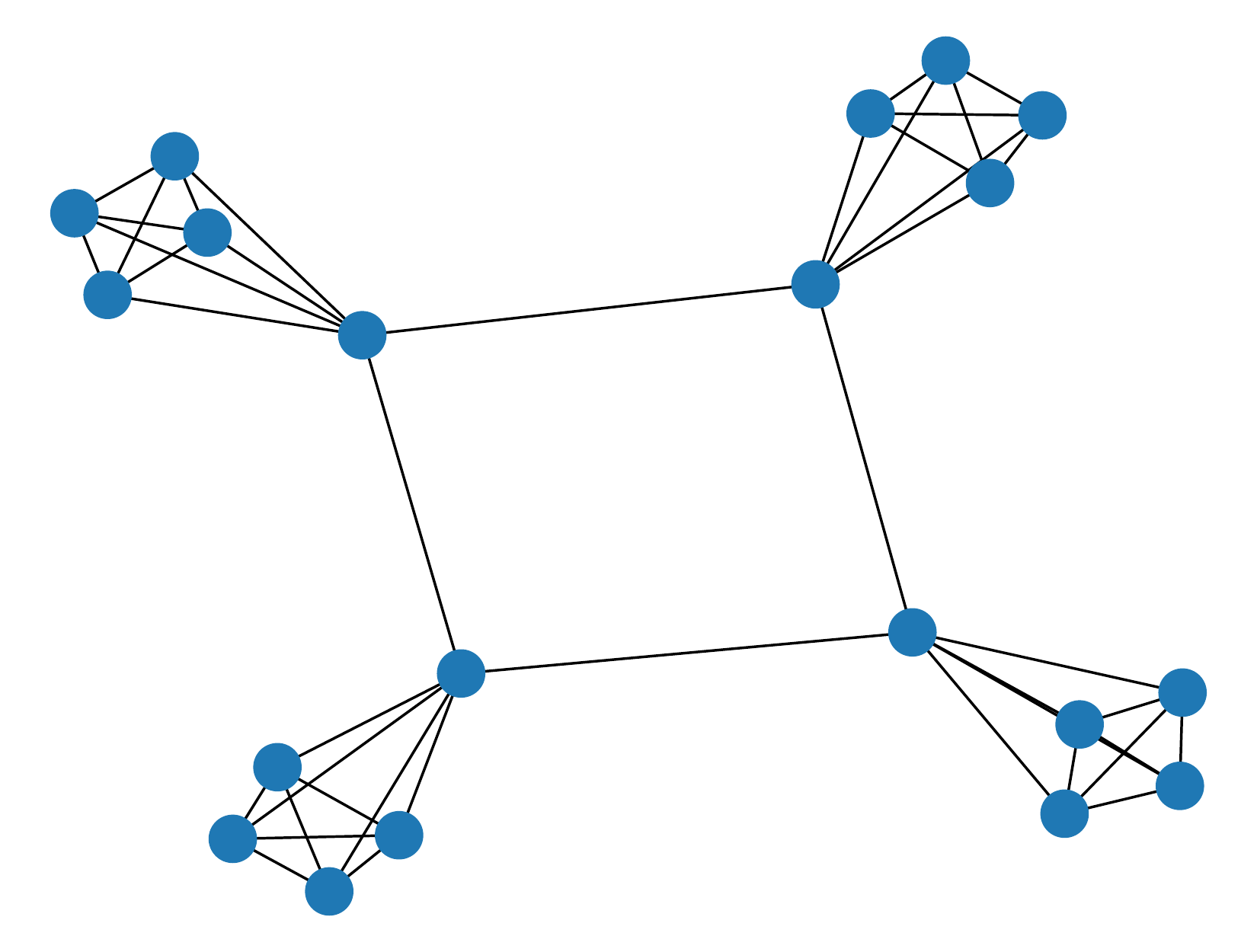}%
  }
  \quad
  \subfloat[game 04\_06\_2]{
    \includegraphics[width=.47\columnwidth]{images-supplementary/game-04_06_1.pdf}%
  }
  \subfloat[game 08\_05\_2]{
    \includegraphics[width=.47\columnwidth]{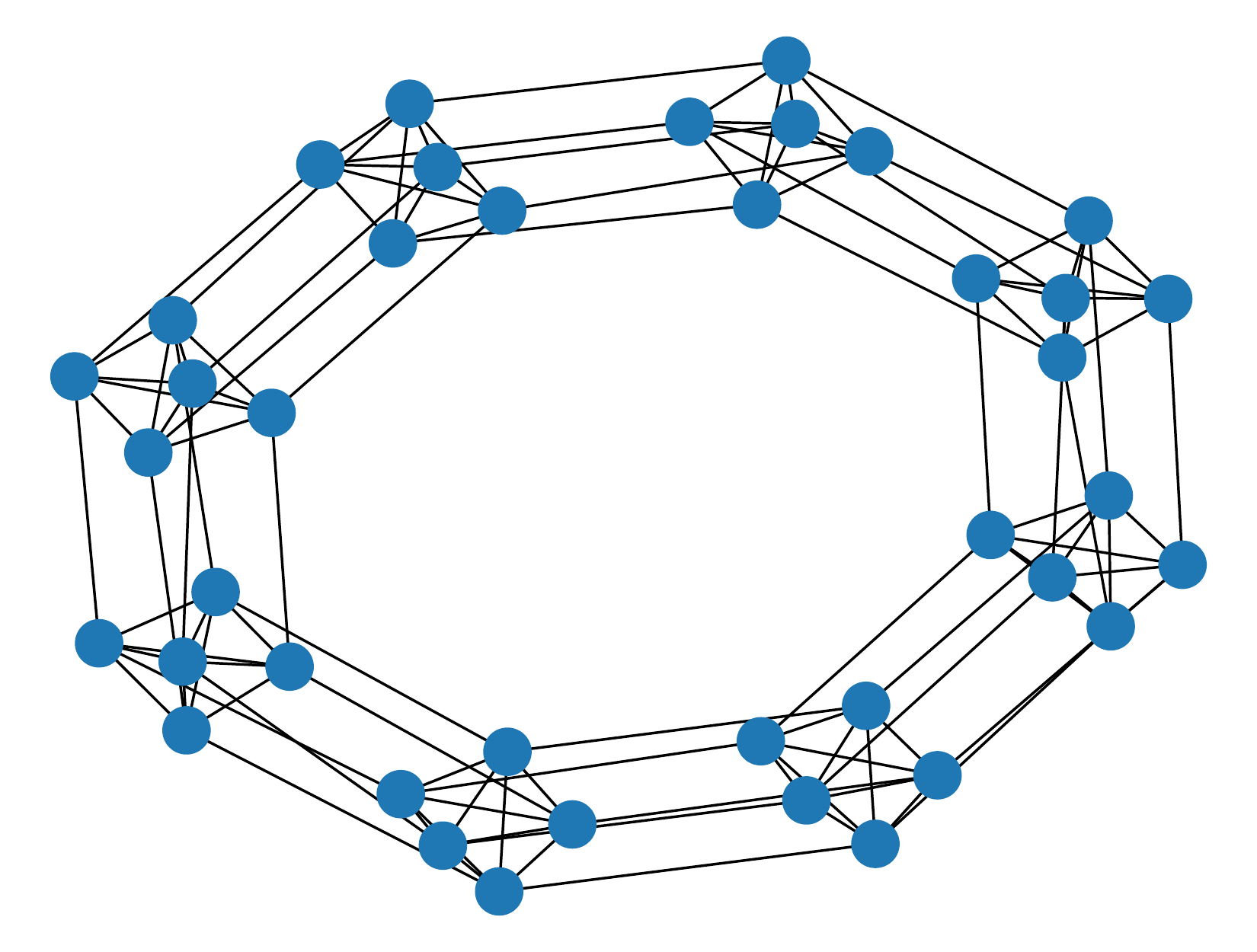}%
  }
  \quad
  \subfloat[game 06\_06\_3]{
    \includegraphics[width=.47\columnwidth]{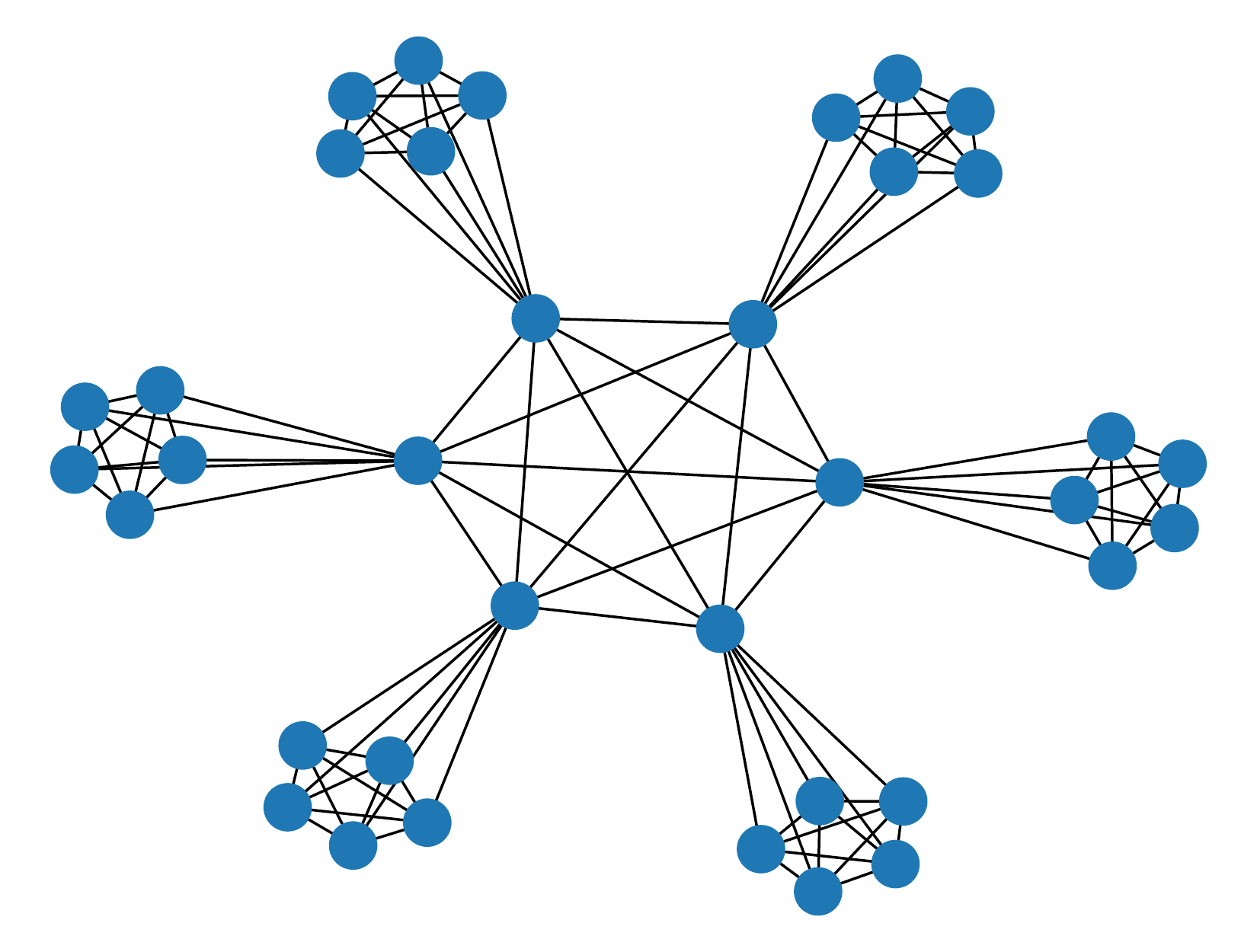}%
  }
  \subfloat[game 10\_05\_3]{
    \includegraphics[width=.47\columnwidth]{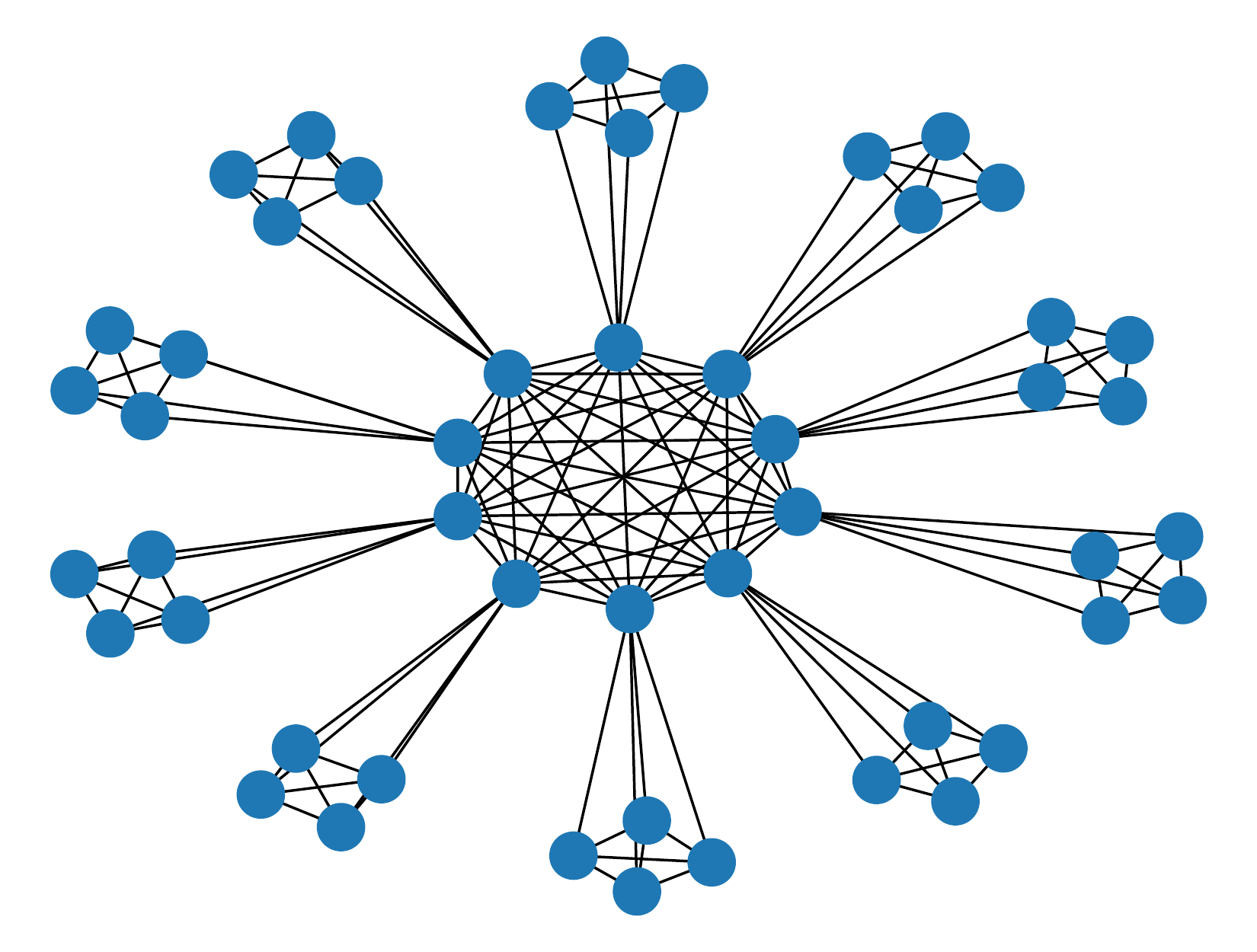}%
  }
  \quad
  \caption{Sample \textit{locally-dense} games from the benchmark set.}%
  \label{fig:sample-games-locally-dense}
\end{figure}

\textit{Locally-dense} graphs were not generated randomly. Each of them was composed of $k$ cliques of size $n$ connected according to the connection rule $r=\{1,2,3\}$:
%There are 3 cliques connection rules:
\begin{itemize}
    \item $r=1$: exactly one vertex from each clique is connected to the two neighboring cliques,
    \item $r=2$: each clique vertex is connected to corresponding vertices from the two neighboring cliques,
    \item $r=3$: exactly one vertex from each clique is connected to all other cliques.
\end{itemize}
Each game is encoded as $k\_n\_r$. For example, $04\_06\_1$ indicates a game based on a graph with 4 cliques of size 6 connected according to the rule 1.
Figure~\ref{fig:sample-games-locally-dense} shows 6 examples of \textit{locally-dense} graphs (2 for each connection rule). 

\emph{Locally-dense} graphs are an important contribution of this paper since previous works (e.g. \cite{bondi2020signal}) did not account for the possibility of different travel times between distinct areas.

In games generation procedure we considered a restricted class of uncertainty matrices, parameterized by $\kappa \in [0, 1]$:
$$
\Pi_{\kappa} = \begin{bmatrix}
1 & \kappa & \frac{\kappa}{2}\\
0 & 1-\kappa & \frac{\kappa}{2}\\
0 & 0 & 1 - \kappa
\end{bmatrix}
$$

%All benchmark games and source code can be downloaded from github.com/easgs/benchmark\_games and github.com/easgs/source\_code.

\section{EASGS parameterization}
EASGS parameters were tuned on a set of $12$ games with $20$ vertices ($3$ games of each type: \textit{sparse}, \textit{moderate}, \textit{dense} and \textit{locally-dense}). The graphs connections and payoffs were generated randomly in order not to bias hyper parameters in any way. All those games were excluded from the final EASGS evaluation tests. 
Table~\ref{tab:parameters} presents all tested parameters and their ranges.
There were $5000$ EASGS parameter tuning runs. In each run, a randomly chosen game and a random selection of parameter values from Table~\ref{tab:parameters} were considered. Afterwards, for each parameter the value with the highest average payoff (calculated across all runs in which this value was selected) was chosen as the final recommendation (bolded in the table).

\begin{table}[ht]
\small
\resizebox{1.0\columnwidth}{!}{%
  \begin{tabular}{l|c|l}
    parameter&symbol&value\\
    \hline
    population size & $n_{pop}$&10, 20, 50, 100, \textbf{200}, 500, 1000, 2000, 5000\\
    \# generations&$n_{gen}$&500, 1000, \textbf{2000}, 5000\\
    \# generations& & \\
    between refreshes&$n_{ref}$&50, 100, 200, \textbf{300}, 500, 1000\\
    mutation rate & $\mathcal{P}_{m}$&0, 0.1, 0.2, 0.3, 0.4, 0.5, 0.6, 0.7, \textbf{0.8}, 0.9, 1\\
    crossover rate & $\mathcal{P}_{c}$&0, 0.1, 0.2, 0.3, 0.4, \textbf{0.5}, 0.6, 0.7, 0.8, 0.9, 1\\
    selection pressure & $\mathcal{P}_{sp}$&0.6, 0.7, 0.8, \textbf{0.9}, 0.95, 1\\
    \# mutation repeats & $m_{limit}$ & 1, 2, 5, $\textbf{10}$, 20, 50\\
  \end{tabular}
 }
 \caption{Parameter values used in the tuning process. Finally recommended values are bolded.}%
\label{tab:parameters}
\end{table}

\begin{figure*}[ht!]
\centering
  \subfloat[population size]{
    \includegraphics[width=0.36\hsize]{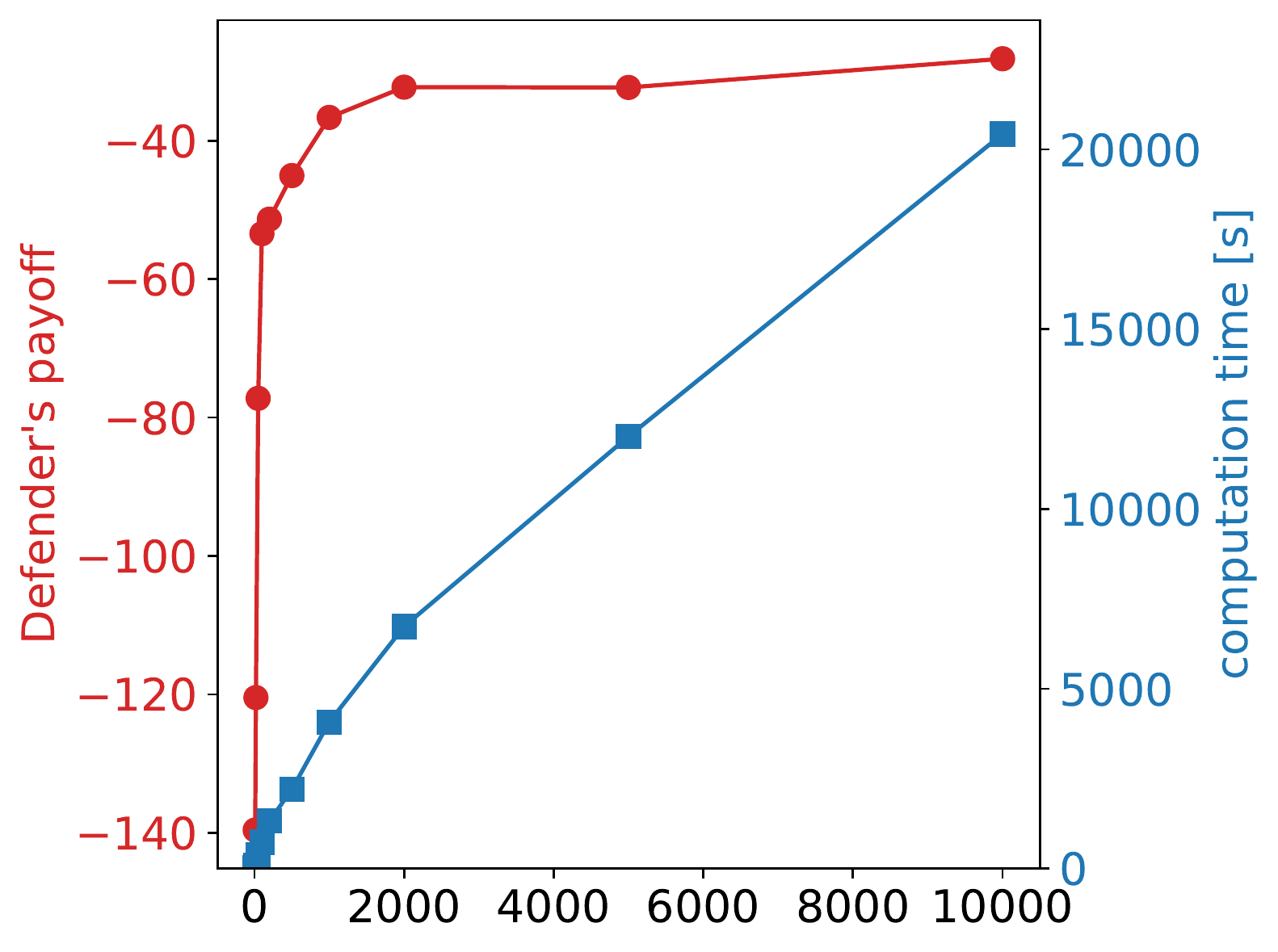}%
    \label{fig:population_size}
  }
  \subfloat[mutation rate]{
    \includegraphics[width=0.36\hsize]{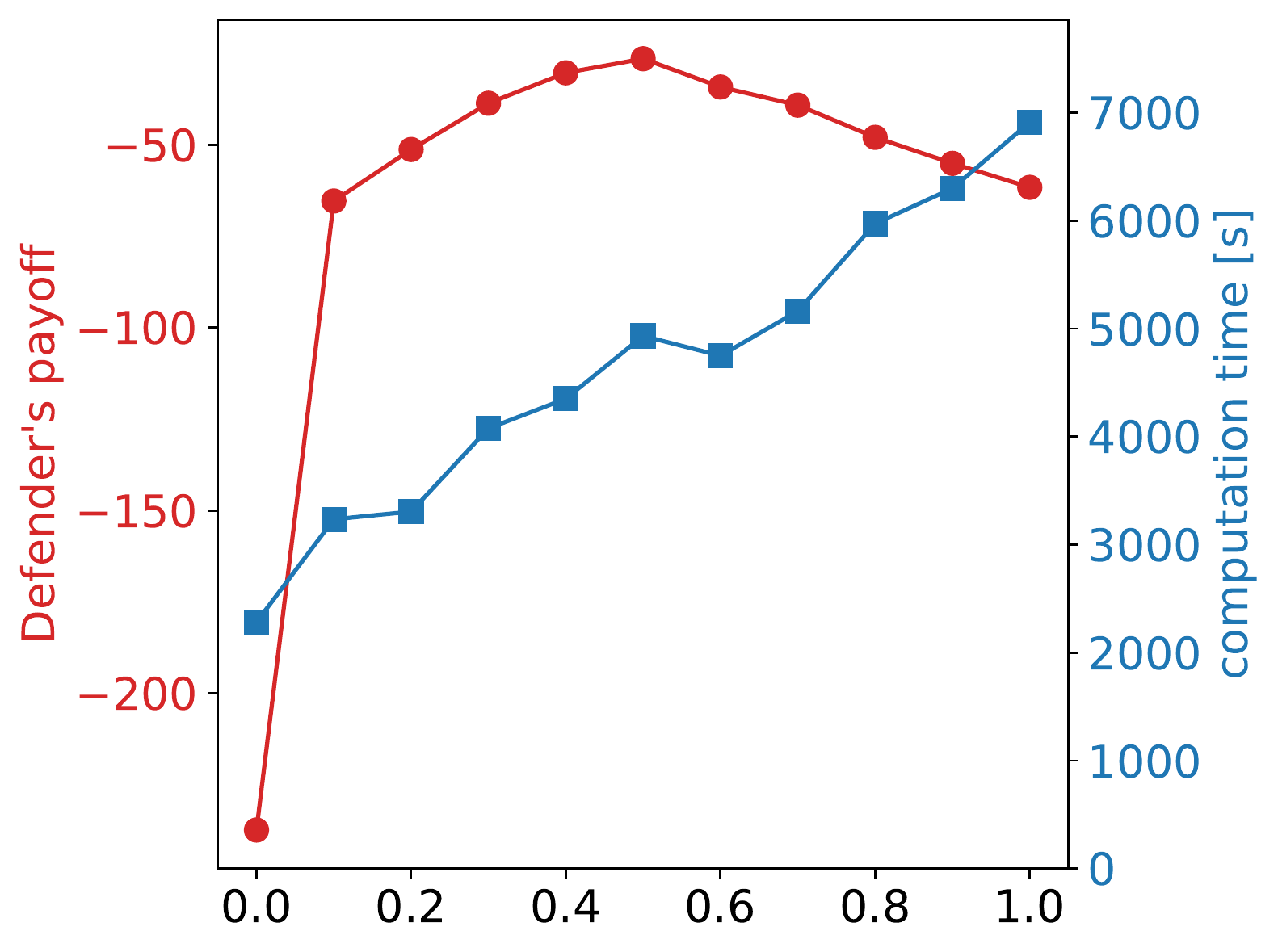}%
    \label{fig:mutation_rate}
  }\\
  
  \subfloat[crossover rate]{
    \includegraphics[width=0.36\hsize]{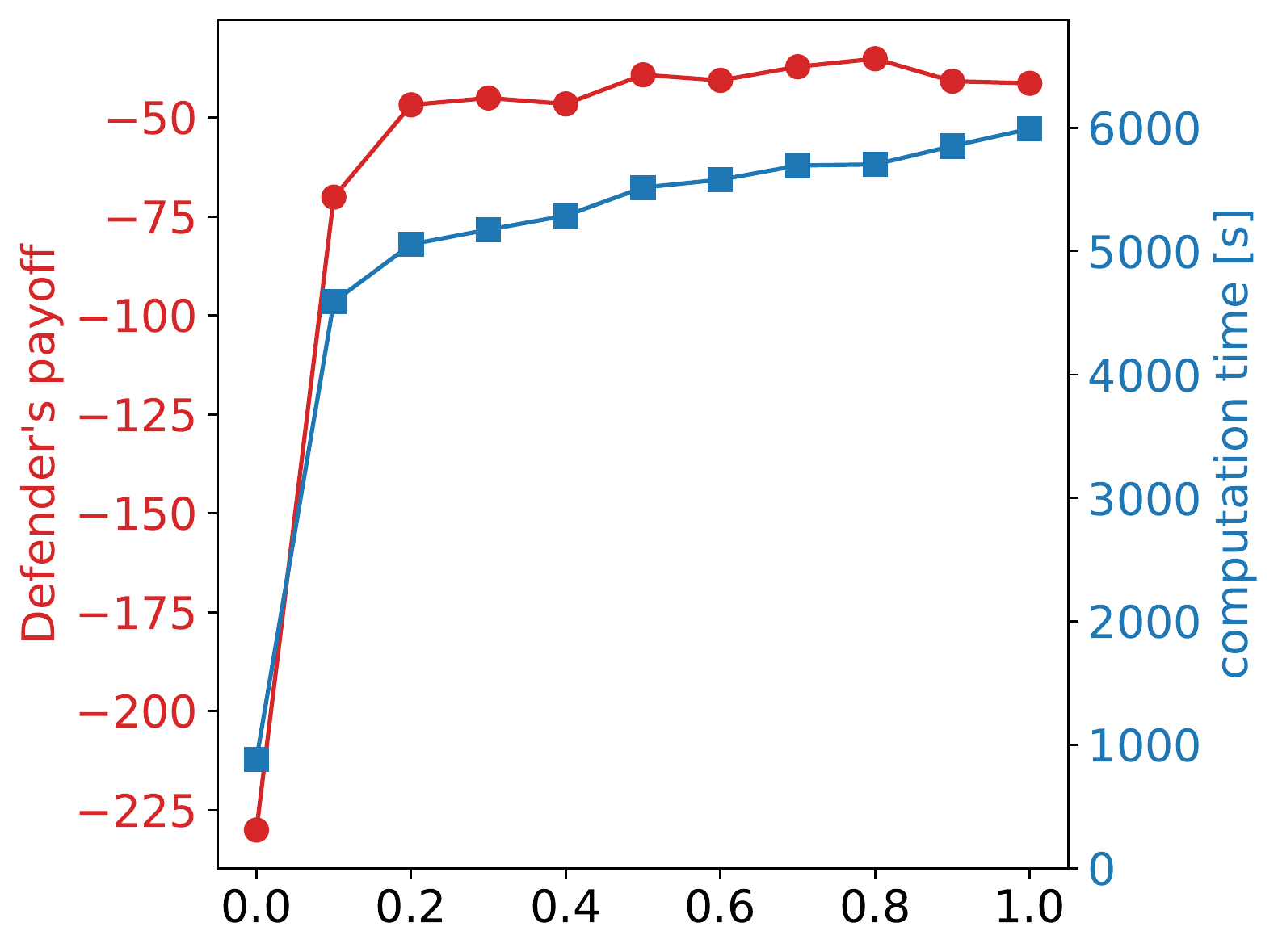}%
    \label{fig:crossover_rate}
  }
  \subfloat[selection pressure]{
    \includegraphics[width=0.36\hsize]{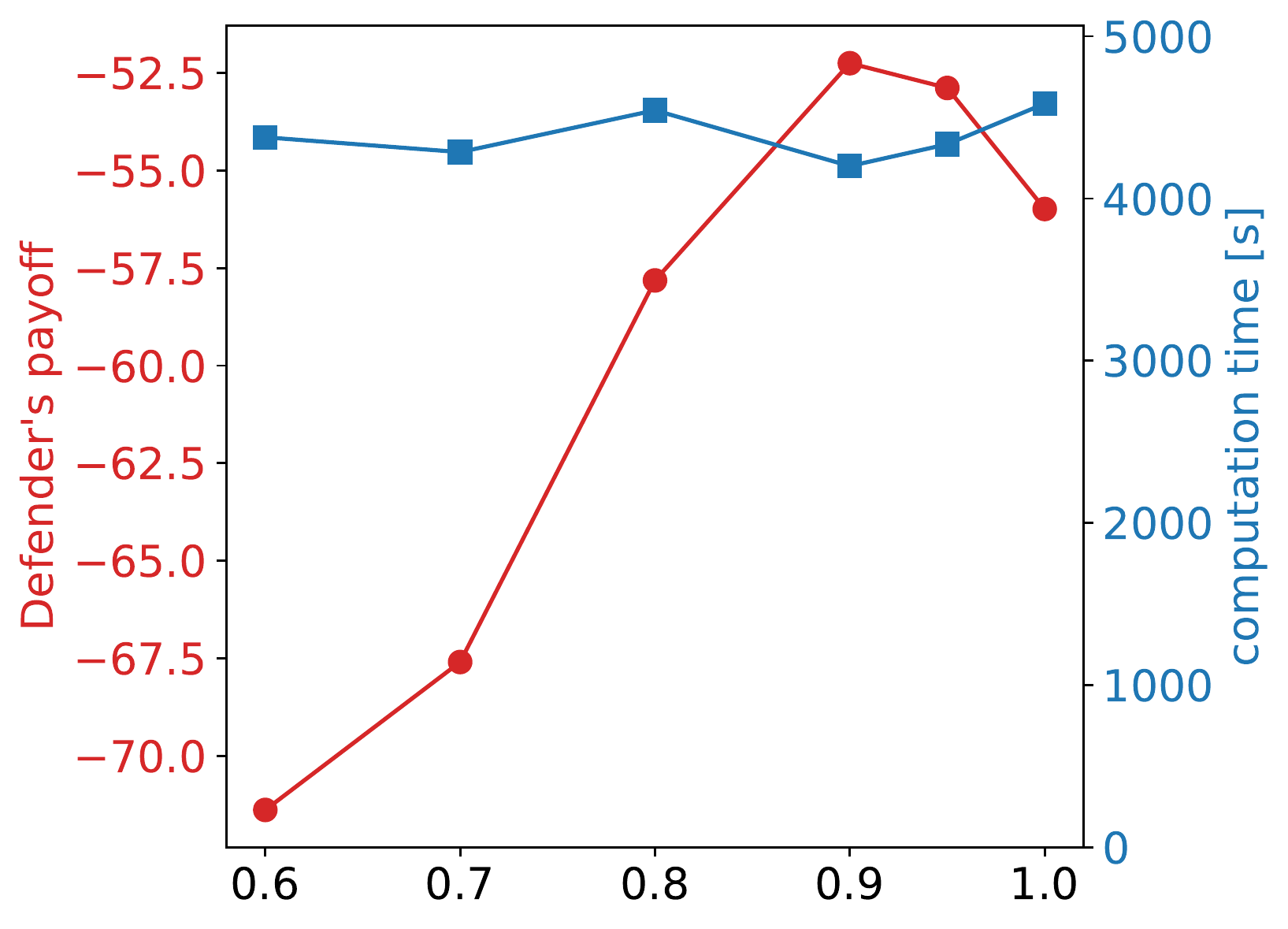}%
    \label{fig:selection_rate}
  }
  \caption{Results of the EASGS parameter tuning phase - the average defender's payoff (circles) and computation time (squares).}
  \label{fig:parameters}
\end{figure*}

Figure~\ref{fig:parameters} shows the average defender's payoff and computation time with respect to the main EASGS steering parameters, calculated in the parameter tuning phase.

Figure~\ref{fig:population_size} presents the results obtained for different \emph{population sizes} $n_{pop}$. The higher the defender's payoffs the more individuals can explore the strategy space and consequently the more potential solutions are considered. For small values, the expected defender’s payoff increases rapidly and then flattens. On the other hand, computation time scales approximately linearly with respect to the population size. Thus, as a trade-off value, $n_{pop} = 200$ was finally chosen.

A relation between the defender's payoff and the \emph{mutation rate} $\mathcal{P}_{m}$ is presented in Figure~\ref{fig:mutation_rate}. Mutation is a key element of the EA process and without it ($p_m=0$) the defender's payoffs are very low. At the same time, too high mutation rate is not a good choice, because in such a situation the majority of strategies are changed randomly in each generation what leads to high fluctuations of the population.
%and good solutions are not preserved enough. 
Computation time slightly grows along with the mutation rate increase due to the increasing number of chromosome modifications. $\mathcal{P}_{m} = 0.8$ was finally selected.

The next investigated parameter was crossover rate $\mathcal{P}_{c}$, presented in Figure~\ref{fig:crossover_rate}. The lack of crossover ($\mathcal{P}_{c} = 0$) results in low payoffs which proves the importance of this operator. No significant differences in terms of payoff and computation time were observed among all other crossover rates. Hence, $\mathcal{P}_{c} = 0.5$ was arbitrary selected.

For the \emph{selection pressure} $\mathcal{P}_{s}$, only values greater than $0.5$ were tested, since $\mathcal{P}_{s} < 0.5$ would mean that lower fitted individuals were more likely to be promoted to the next generation. Computation time remains approximately constant across all tested values. The best results (highest defender's payoffs) were obtained for $\mathcal{P}_{s}=0.9$ (cf. Figure~\ref{fig:selection_rate}).

\section{Strategy visualization}

\begin{figure*}
\centering
\includegraphics[width=.7\textwidth]{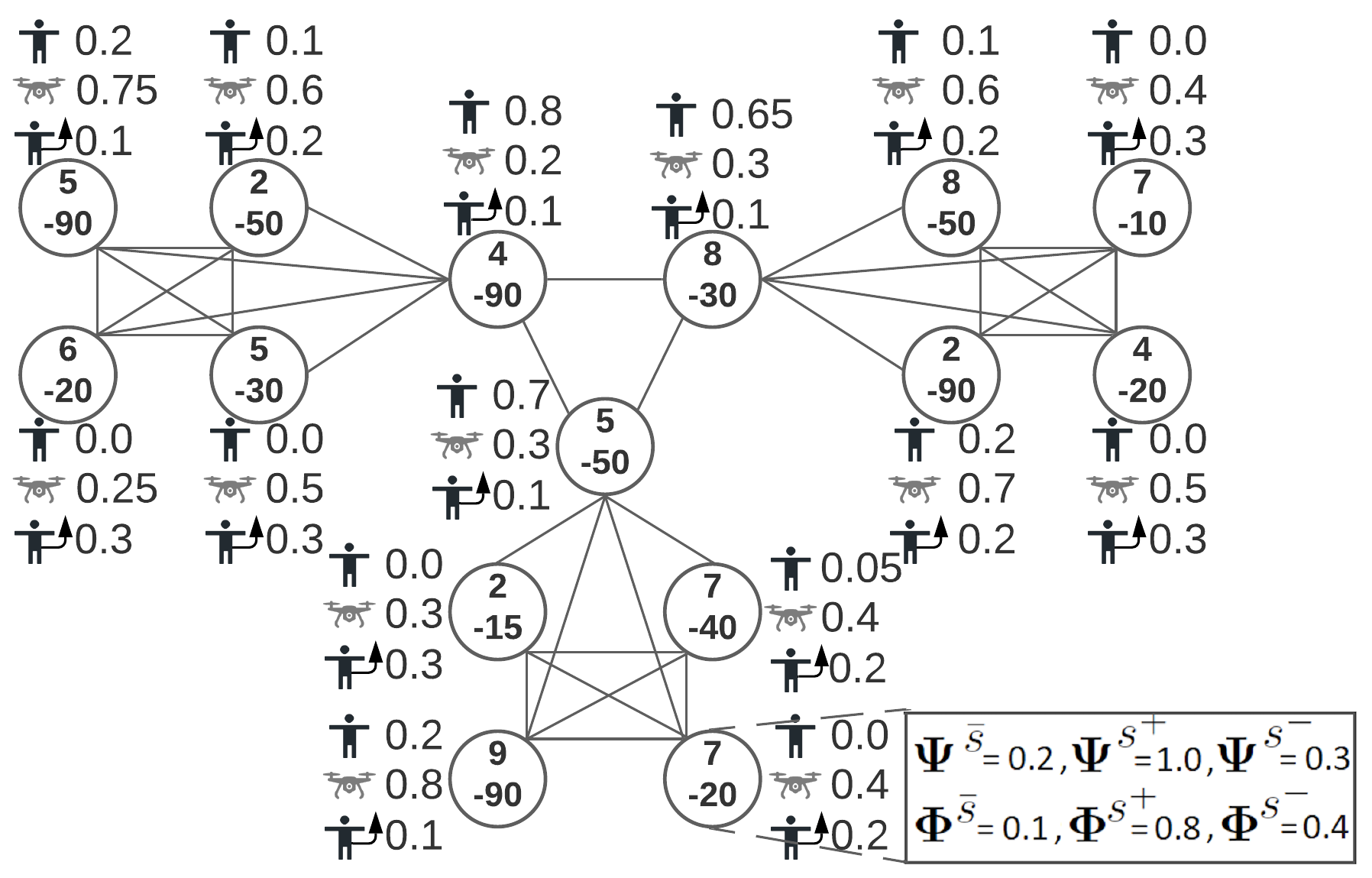}%
\caption{Visualization of a sample output strategy generated by EASGS. Numbers in each vertex indicate the adversary's reward (top) and the defender's penalty (bottom) in case of a successful attack on this target. Next to each target the following coverage probabilities are presented (from top to bottom): patrollers' allocation, sensors' allocation, and patrollers' reallocation. For the sake of clarity of the figure the signalling strategies in the targets are omitted, except for one target located in the bottom-right part of the figure.}
\label{fig:SGS_game}
\end{figure*}

Figure~\ref{fig:SGS_game} presents 
%visualisation of 
a strategy generated by EASGS for a sample \textit{locally-dense} game with 15 vertices (targets). The defender has 3 patrollers and 7 sensors (drones) at their disposal. Numbers in each vertex indicate the attacker's reward (top) and the defender's penalty (bottom) in case of a successful attack on a given target. Next to each target the following coverage probabilities are presented (from top to bottom): patrollers' allocation, sensors' allocation, and patrollers' reallocation. Please note that these probabilities are not encoded directly in a chromosome and are computed based on the list of allocation pure strategies and their corresponding probabilities, denoted by $e$ and $q$, resp. in the chromosome definition. For the sake of clarity, the sensors' signaling strategy is presented only for one target.
Please note, that the three central vertices (connected to 5 targets) are more likely to have patrollers allocated. Intuitively, placing patrollers in highly connected vertices opens more possibilities in the reaction stage, i.e. a patroller can move to any of the connected targets in case of sensor's signaling. Please observe, that presented coverage probabilities are not strictly proportional to the target's utilities since the expected payoff depends also on the signaling strategy, patrollers' reallocation, attacker's reaction strategy to signaling, and uncertainty (detection and observational).

Sensor's signaling strategy, denoted by $\boldsymbol{\Psi^\theta}$ and $\boldsymbol{\Phi^\theta}$, determines how likely it is that the sensor allocated in a given target will send signal $\sigma_0$ in state $\theta \in \{\bar{s}, s^+, s^-\}$. Please note, that the signal sent out by the sensor is either $\sigma_0$ or $\sigma_1$, therefore the probabilities of sending signal $\sigma_1$ are equal to $1-\boldsymbol{\Psi^\theta}$ and $1-\boldsymbol{\Phi^\theta}$, respectively. Usually, the probability of sending $\sigma_1$ (strong signal) is low in state $s^+$ (there is a patroller who will visit the target in the reaction stage) because in such case strong signal may alarm the attacker. On the contrary, sending $\sigma_1$ in state $\bar{s}$ is desirable since there is no patroller in the neighbourhood, the attacker cannot be caught and the only chance is to scare them and provoke to flee.

\section{EASGS operators - ablation study}
In EASGS several novel components compared to standard EA realization are implemented, including three types of mutation, mutation repetition, the removal of strategies in crossover, specific population \textit{refreshing}, and local coverage optimization.
In order to better understand the role of particular enhancements and their actual impact on the EASGS performance, several ablation experiments were executed in which 9 different variants of EASGS were tested. In each of them, one EASGS component was removed or replaced with an alternative, and the rest of the method remained unchanged. These variants are defined in the following way:
\begin{itemize}
    \item $EASGS$-$C$ - crossover operator is not applied,
    \item $EASGS$-$M$ - none of mutation operators is applied,
    \item $EASGS$-$M_z$ - mutation of type $M_z$ is not applied, $z \in \{1,2,3\}$,
    \item $EASGS$-$LO$ - local optimization procedure is not applied,
    \item $EASGS$-$R$ - population \textit{refreshing} in case population stagnation is not applied,
    \item $EASGS$-$C_r$ - procedure of removing randomly selected pure strategies in crossover operator is not applied,
    \item $EASGS$+$C_o$ - crossover operator ($C_o$), originally proposed in~\cite{ZychowskiMandziuk2021}, is applied, instead of the baseline EASGS crossover.
\end{itemize}

Table~\ref{tab:results_removal} presents a comparison of all the above variants with the baseline version of EASGS in terms of the mean defender's payoff.
\begin{table*}[ht]
\setlength\tabcolsep{4px}
\resizebox{1.0\textwidth}{!}{
  \begin{tabular}{c|c|c|c|c|c|c|c|c|c|c}
 & $EASGS$ & $EASGS$-$C$ & $EASGS$-$M$ & $EASGS$-$M_1$ & $EASGS$-$M_2$ & $EASGS$-$M_3$ & $EASGS$-$LO$ & $EASGS$-$R$ & $EASGS$-$C_R$ & $EASGS$+$C_o$ \\\hline
\textbf{sparse} & \textbf{-91.32} & -232.25 & -132.98 & -112.28 & -105.64 & -98.64 & -95.10 & -96.84 & -121.44 & -92.14 \\
\textbf{moderate} & \textbf{-69.92} & -168.56 & -119.08 & -88.45  & -83.12  & -79.70 & -73.53 & -75.77 & -105.79 & -70.34 \\
\textbf{dense} & \textbf{-51.47} & -178.23 & -85.42  & -62.96  & -59.98  & -55.76 & -54.03 & -56.24 & -77.24  & -51.97 \\
\textbf{locally-dense} & \textbf{-54.36} & -150.92 & -85.51  & -67.22  & -63.72  & -59.67 & -57.02 & -57.76 & -78.06  & -54.88
  \end{tabular}
}
\caption{Comparison of average defender's payoff in various EASGS setups.}
\label{tab:results_removal}
\end{table*}
Each of the above modifications %(i.e. removing of any of the above-listed EASGS components)
leads to worsening the average payoff compared to the baseline EASGS formulation, for each game type, which confirms that all EASGS components are relevant. As could be expected, the worst results were obtained after removing one of the key evolutionary operators - crossover ($EASGS$-$C$) or mutation ($EASGS$-$M$).

\subsection{$EASGS$-$C$}
Crossover merges and mixes strategies from two chromosomes and its removal leads to the situation where all encoded strategies remain pure, which is ineffective in most of the cases. 

\subsection{$EASGS$-$M$}
%As far as the mutation removal is concerned, observe that 
The role of mutation is to explore new solutions by modifying pure strategies in the chromosomes. When mutation is disabled, all pure strategies drawn for the initial population remain unchanged throughout all generations. The crossover operator will mix them, but no new pure strategy will be created/added as a result.

\subsection{$EASGS$-$M_1$}
Looking at the relevance of particular mutation types, the most important mutation is $M_1$. Its removal ($EASGS$-$M_1$) causes the biggest deterioration of results compared to removing other mutations. Without $M_1$ the algorithm is unable to \textit{freely} change probabilities of pure strategies in the chromosome. They can only be modified by the crossover operator, albeit in a quite restricted manner.
%but the crossover always halves them and does not adjust to the expected payoff.

\subsection{$EASGS$-$M_2$}
Resigning from $M_2$ mutation is less harmful for the quality of results but, at the same time, thanks to the application of $M_2$, the algorithm is in principle able to generate any arbitrary pure strategy, thus converge to the optimal selection of pure strategies in a chromosome. Removing this mutation would mean that the sole operator that changes the set of pure strategies in a chromosome (patrollers allocation, sensors allocation) would be $M_3$, whose application changes the allocation of the target coverage to resources in a greedy manner. This, however, does not necessarily increase the expected payoff. Moreover, $M_2$ is the only operation which changes the defender's signaling strategy, so removing it would render the signaling strategy fixed during the whole EASGS run.
%(drawn at the population initialization stage). 
%and does it in a greedy way (increase coverage chosen by the adversary).

\subsection{$EASGS$-$M_3$}
Removal of $M_3$ mutation ($EASGS$-$M_3$) has the lowest impact. This mutation modifies a given pure strategy by adding targets to patrollers'/sensors' allocation strategy. The same operation can be performed with $M_2$, so $M_3$ is potentially redundant, however its addition improves EASGS convergence speed. Instead of changing pure strategies randomly (as $M_2$ does) and \textit{luckily} finding the modification that improves the result, $M_3$ often makes the right choice in just one mutation application. 
%Consequently, while it  does not strongly influence the quality of the final result it improves the convergence.
%, nonetheless enables the algorithm to move individuals to more promising search space areas faster. 
The average number of generations required for finding the best solution was equal to $462$ when $M_3$ was disabled, compared to $126$ iterations when this mutation was enabled.

\subsection{$EASGS$-$LO$}
Relatively good results were obtained by $EASGS$-$LO$. Local optimization ensures that in the generated strategies 
%\textit{unleash the potential} i.e. 
no two patrollers are allocated to the same vertex and that reallocation node is connected to the patroller's current (allocated) position. Removing local optimization phase results in strategies which do not fulfill the above conditions, e.g., some patrollers may not be able to reallocate to the vertex encoded in their reallocation strategy. Thus, fewer targets are covered, and the defender's payoff decreases. EASGS is able to recover from this chromosome non-optimality caused by disabling the local memetic optimization phase, but the number of generations required to converge to a good solution visibly increases, on average, from $126$ generations in the case of $EASGS$ to $576$ generations for $EASGS$-$LO$. We hypothesize that these infeasible strategies, because of their non-optimality in terms of allocation, are lower-fitted and less probable of being promoted to the next generation in the selection procedure. Consequently, $EASGS$-$LO$ payoffs do not deviate much from $EASGS$ ones.

\subsection{$EASGS$-$R$} 
\begin{figure}[ht]
\centering
  \subfloat[$EASGS$]{
    \includegraphics[width=0.98\columnwidth]{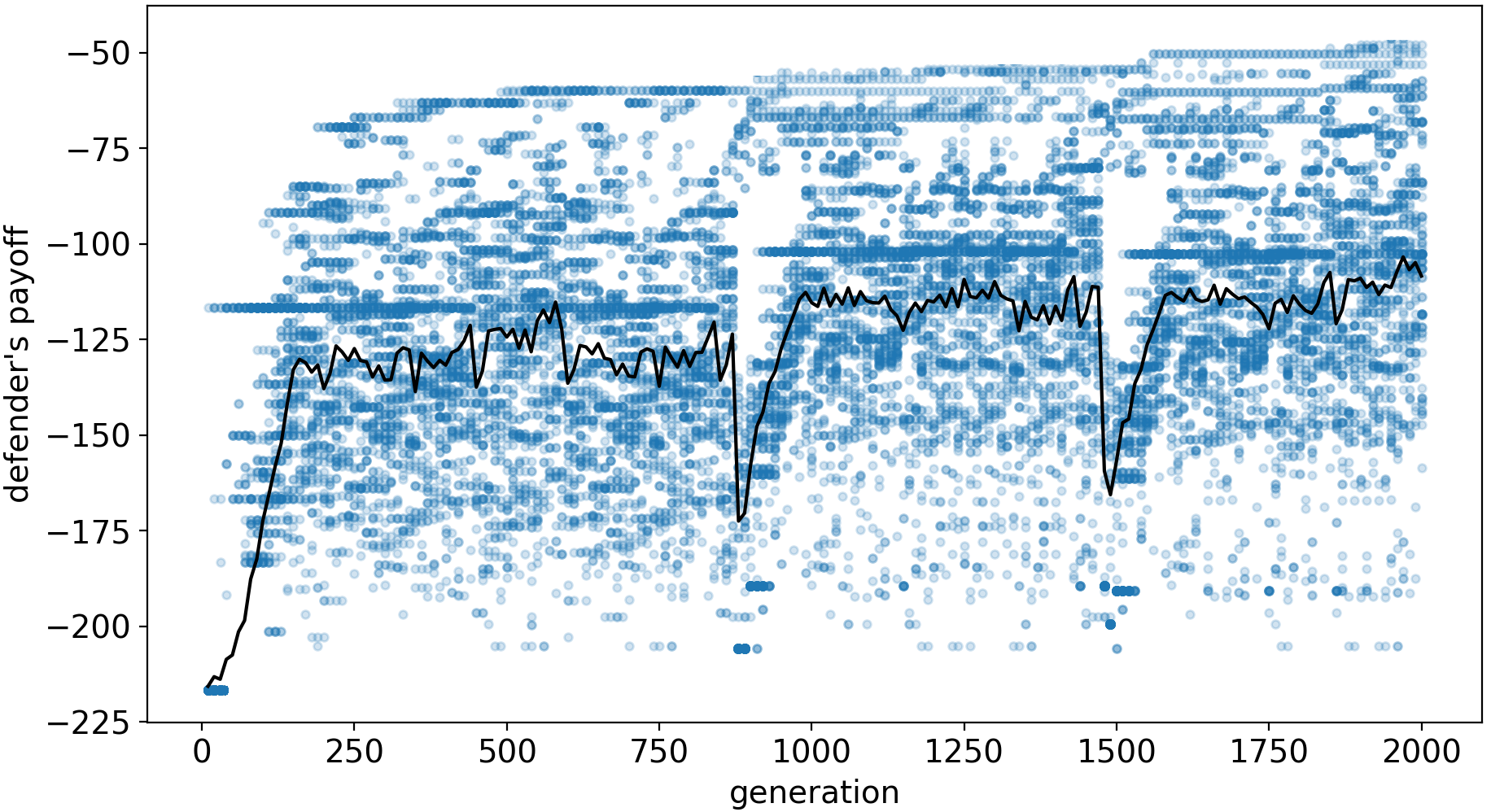}%
  }\\
  \quad
  \subfloat[$EASGS$-$R$]{
    \includegraphics[width=0.98\columnwidth]{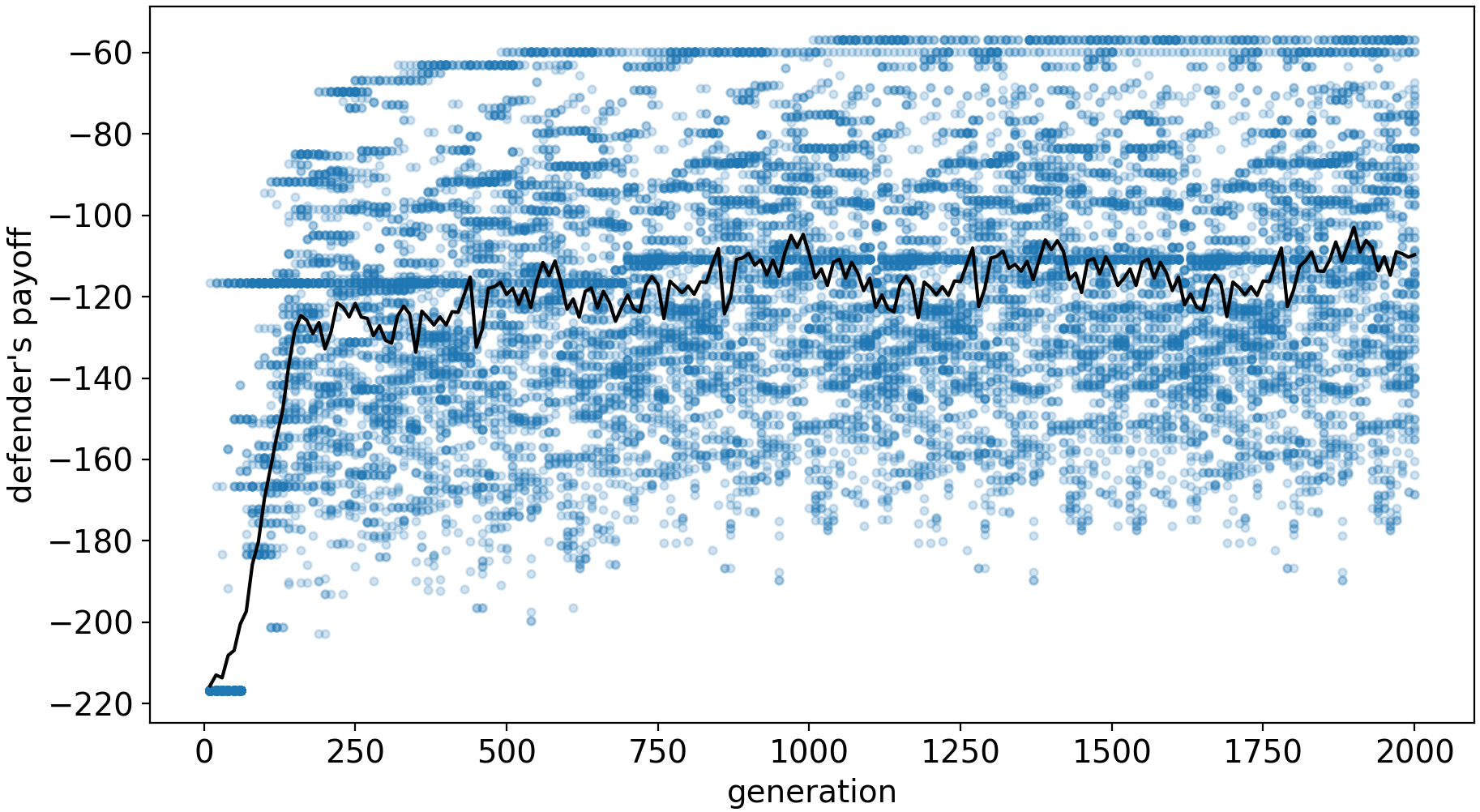}%
  }
  \caption{A typical convergence pattern of the baseline algorithm ($EASGS$) vs its version without population refreshing ($EASGS$-$R$). Blue dots represent individual chromosomes' fitness values. Dark blue areas represent more individuals with a particular fitness value. The solid black line is the average defender's payoff across the population. In Figure (a) two \textit{refreshing} points are clearly visible. Their application causes a temporary drop in the average payoff (note that more positive denotes a better payoff), but after a few generations, the population recovers at a level slightly higher than that immediately before refreshing. In both figures, we see the best chromosomes continuing to increase in payoff value.}%
  \label{fig:convergence}
\end{figure}

Running the proposed algorithm without the ability to \textit{refresh} the population ($EASGS$-$R$) worsens the results by about 6\% on average. Refreshing breaks the population stagnancy and pushes it to another area of the solution space. Figure~\ref{fig:convergence} compares a   population  convergence with and without the refreshing procedure. It can be observed that although refreshing temporarily decreases the average population payoff, the subsequent application of other operators (crossover and mutation) leads to creation of better-fitted individuals.

\subsection{$EASGS$-$C_r$} 
The next tested component, removal of randomly selected pure strategies in the crossover operator, is designed to avoid generation of chromosomes with a large number of pure strategies with tiny probabilities. Each crossover operation merges all pure strategies from both child chromosomes. Thus, without the removal procedure, after a few generations, chromosomes would become large. We observed that after 50 generations some of the probabilities in mixed strategies were less than $10^{-5}$ and had nearly no influence on the final result. In such cases, the mutation performed on such a pure strategy barely impacted the individual's fitness, and population evolution decelerated. Hence, disabling the removal procedure ($EASGS$-$C_r$) results in significantly lower payoffs. Moreover, calculating the payoff based on a large number of such pure strategies is very time-consuming. $EASGS$-$C_r$ computation time was almost an order of magnitude bigger than that of $EASGS$: 13.4 minutes for $EASGS$-$C_r$ versus 1.5 minutes for $EASGS$, on average.

\subsection{\textit{EASGS+C\textsubscript{o}}}
We compared the efficacy of our crossover operator with the one ($C_o$) proposed in~\cite{ZychowskiMandziuk2021}. The two operators differ by the way in which probabilities are assigned to pure strategies in the resulting child chromosome. In the crossover proposed in this paper, probabilities are chosen proportionally to the 
%distribution based on defenndein $C_o$ crossover yoff sopure strategy. %Generally, the greater payoff of the pure sgiven trategy is, the more contribution in the result m. the greater probabiliixed strategy it has, i.ety of this strategy is assigned. 
the expected defender's payoff (when the respective pure strategy is played). In $C_o$ crossover, a probability of each pure strategy in the resulting child chromosome equals half of the respective probability in the parent chromosome. If a given pure strategy exists in both parent chromosomes, it is listed only once in the child chromosome, with mean probability.

%Formally, sensors' signaling strategies $\boldsymbol{\Psi^\theta}$ and $\boldsymbol{\Phi^\theta}$ in the child chromosome contain  the averaged signaling probability from each parent.  
%A result of the crossover operation on chromosomes $CH_1$ and $CH_2$ will be the new chromosome, $CH_{1\text{-}2}$:
%\begin{multline*}
%CH_{1\text{-}2}=\{(e^1_1,\frac{q^1_1}{2}),(e^1_2,\frac{q^1_2}{2}), \ldots, (e^1_{d_1},\frac{q^1_{d_1}}{2}),\\(e^2_1,\frac{q^2_1}{2}),(e^2_2,\frac{q^2_2}{2}), \ldots, (e^2_{d_2},\frac{q^2_{d_2}}{2}), \boldsymbol{\Psi^\theta_{1\text{-}2}}, \boldsymbol{\Phi^\theta_{1\text{-}2}}\},
%\end{multline*}
%
%where\\
%\smallskip
%$\boldsymbol{\Psi^\theta_{1\text{-}2}} = [\frac{1}{2}(\Psi^\theta_{1,1} + \Psi^\theta_{2,1}), % \frac{1}{2}(\Psi^\theta_{1,2} + \Psi^\theta_{2,2}), 
%\ldots,
%\frac{1}{2}(\Psi^\theta_{1,\mathcal{N}} + \Psi^\theta_{2,\mathcal{N}})]$, \\
%$\boldsymbol{\Phi^\theta_{1\text{-}2}} = [\frac{1}{2}(\Phi^\theta_{1,1} + \Phi^\theta_{2,1}), % \frac{1}{2}(\Phi^\theta_{1,2} + \Phi^\theta_{2,2}), 
%\ldots,
%\frac{1}{2}(\Phi^\theta_{1,\mathcal{N}} + \Phi^\theta_{2,\mathcal{N}})]$.
%\smallskip
Table~\ref{tab:results_removal} shows that the newly-proposed crossover slightly outperforms $C_o$ in terms of the average defender's payoff. However, the true strength of our crossover lies in the algorithm's convergence speed. The average number of generations before an individual--finally returned as a result--is found for the first time, decreases from $1060$ in $EASGS$+$C_o$ to $750$ in $EASGS$.

\subsection{Erdos-Renyi graphs}
\textit{Sparse}, \textit{moderate} and \textit{dense} games are defined on game graphs generated according to Watts–Strogatz model \cite{watts1998collective}. In order to check if EASGS is not biased towards this particular type of graphs we prepared a set of 18 (very different from Watts–Strogatz) instances of Erdos-Renyi graphs~\cite{erdos1960evolution} and evaluated all methods on these graphs, using the same steering parameters as in the main experiment.

\begin{table}[ht]
\small
\setlength\tabcolsep{4px}
\resizebox{1.0\columnwidth}{!}{
  \begin{tabular}{c|c|c|c|c}
 & SBP & SBP+W & m-CombSGPO & EASGS \\\hline
\textit{sparse} & -91.91 (67\%) & \textbf{-90.32 (83\%)} & -365.82 (0\%) & -92.57 (17\%) \\
\textit{moderate} & -78.51 (0\%) & -76.84 (33\%) & -231.39 (0\%) & \textbf{-74.58 (67\%)} \\
\textit{dense} & -68.34 (0\%) & -67.41 (17\%) & -141.06 (0\%) & \textbf{-62.11 (83\%)} \\
  \end{tabular}
}
\caption{The average defender's payoff for Erdos-Renyi graphs.}
\label{tab:erdos-renyi-graphs}
\end{table}

The results presented in Table~\ref{tab:erdos-renyi-graphs} are qualitatively similar to those reported in the main paper (Table 2), which supports the claim that parameter tuning is not biased towards specific game graphs.

\end{document}